\documentclass[3p]{elsarticle}

\usepackage{amsmath}
\usepackage{amssymb}

\journal{Physica A}

\newcommand{\dd}{\mathrm{d} }
\newcommand{\Ip}{\mathrm{I}}

\newcommand{\possessivecite}[1]{\citeauthor{#1}'s \citep{#1}}

\usepackage{url}

\begin{document}
\begin{frontmatter}

\title{The Concept of  Force in Population Dynamics }
\date{DOI: 10.1016/j.physa.2019.121736}

\author{John Hayward\corref{cor1}}
\cortext[cor1]{Corresponding author}
\ead{john.hayward@southwales.ac.uk}

\author{Paul A. Roach\corref{cor2}}

\address{School of Computing and Mathematics, University of South Wales, Pontypridd, CF37 1DL, Wales, UK}

\begin{abstract}
The area of population dynamics has a rich history of the development and analysis of models of biological and social phenomena using ordinary differential equations. This paper describes a method for understanding the influence one variable exerts on another in such models as a force, with the relative effects of these forces providing a narrative explanation of the curvature in variable behaviour. Using the stock/flow form of a model, a symbolic notation is developed that identifies the forces with the causal pathways of the model's feedback loops. A force is measured by its impact, defined as the ratio of acceleration to rate of change, computed by differentiation along its associated pathway between variables. Different phases of force dominance are determined to enhance the standard stability analysis of the models, providing an explanation of model behaviour in Newtonian mechanical terms. The concepts developed are applied to well-known models from mathematical biology: the Spruce Budworm model, where force dominance identifies scenarios that give clarity to intervention points; and the Lotka-Volterra predator-prey model where the analysis highlights the importance of dissipative forces in achieving stability. Conclusions are drawn on the explanatory power of this approach, with suggestions made for future work.

\end{abstract}


\end{frontmatter}

\section{Introduction}

Sociophysics is a field of science that seeks to understand the behaviour of humans, and other agents, using the theories and techniques of physics such as statistical mechanics and thermodynamics \citep{chakrabarti2007econophysics,castellano2009statistical,galam2012sociophysics,schweitzer2018sociophysics,kutner2018econophysics}.
One common technique is the construction of systems of ordinary differential equations (ODEs), an approach which has long been successful in modelling population dynamics in human, biological and social systems. Examples of application areas include interacting species  \citep{murray2002mathematical,freedman1980deterministic,turchin2003complex} and the spread of disease \citep{kermack1927contribution,anderson1992infectious},  with many standard models being applied to social modelling, such as conflict \citep{lanchester1916aircraft,burbeck1978dynamics,vitanov2010verhulst,vitanov2012discrete}, geopolitics \citep{richardson1960arms,turchin2003historical}, interacting agents \citep{caram2010dynamic,caram2015cooperative} and  social diffusion, for example  \citep{coleman1964introduction, bass1969new,hayward2005general,bettencourt2006power,zhao2011sihr, ausloos2012econophysics, jeffs2016activist}.    Models are constructed by making assumptions concerning  dynamical processes, such as births, predation and infection, which depend on the state variables. In turn, the rates of change of those variables depend  on the contribution of the dynamical  processes.   Thus a  model, with $n$ state variables $x_i$:
\begin{equation}
\frac{\dd x_i}{\dd t} =f_i(x_1,\dots, x_j,\dots,x_n)\triangleq f_i(x_j) \mbox{, \hspace{0.2cm}  }  i,j = 1, \dots, n \label{ode.eq}
\end{equation}
contains a network of dependencies between state variables, representing cause and effect, derived from the model's dynamical processes.  Models are analysed  analytically  and numerically,  with the results interpreted in terms of  the model assumptions. However, the model equations (\ref{ode.eq}) do not explicitly encapsulate the effects of the assumptions on variable behaviour, often due to algebraic simplifications in equation presentation, thus determining the contribution of each  dynamic process to variable behaviour is not normally possible.

By comparison the system dynamics methodology, pioneered by JW Forrester \citep{forrester1961industrial,forrester1968principles,sterman2000business}, encapsulates the dynamic assumptions in network form. Although the models are usually presented in a diagrammatic notation, where the cause and effect of the   dynamic processes are explicit, the diagram represents all the model equations, differential and algebraic, prior to any simplification \citep{sterman2000business}. This system dynamics methodology has been successfully used in, for example, business, environmental and social modelling, where model results are interpreted using the feedback loops made explicit by the model diagrams. Although models are normally analysed with computer simulation, system dynamics models can be reduced to differential equations and analysed analytically. However,  the causal structure is then lost, as explained above.

\begin{figure}[!ht]
          \begin{center}
   \includegraphics[height=3.5cm] {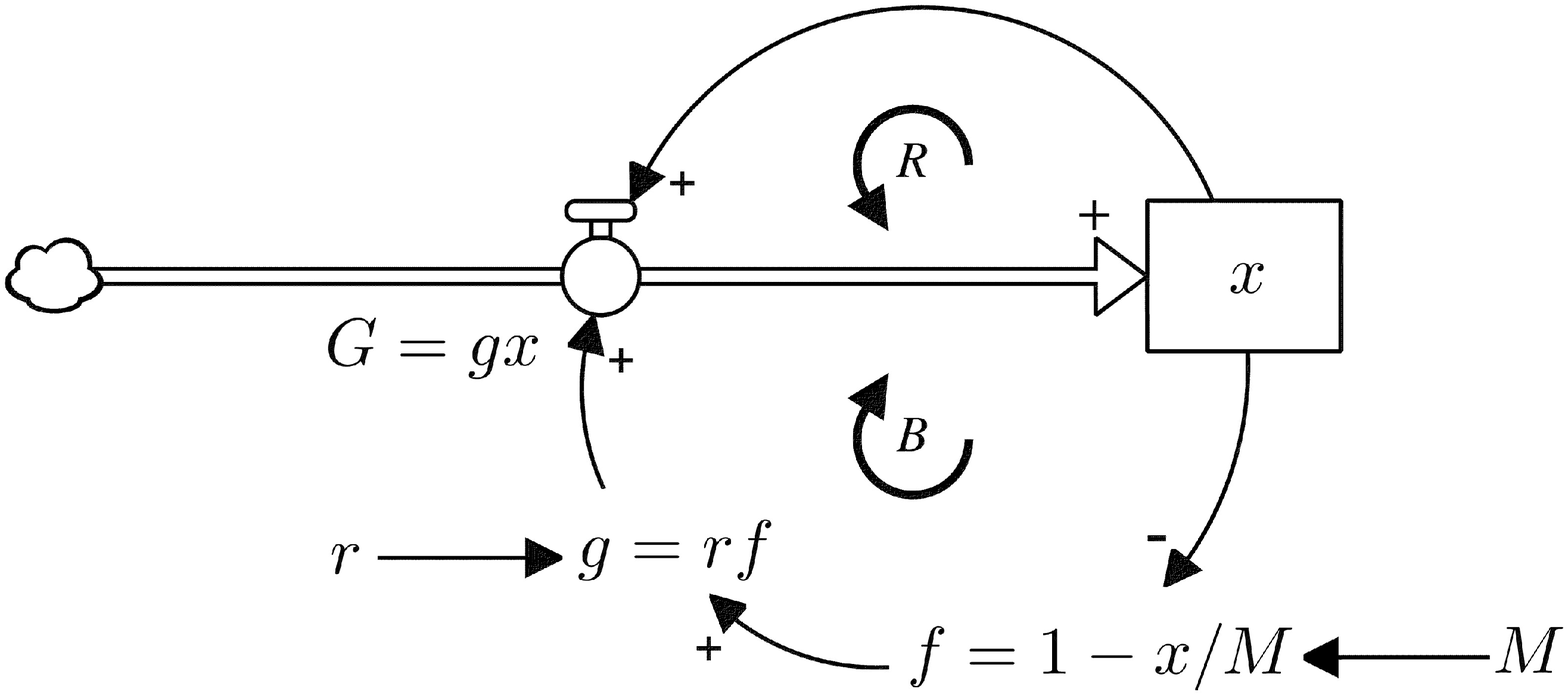}
       \end{center}
    \vspace{-15 pt}
    \caption{\small{Verhulst model for population $x$ in system dynamics notation, with feedback loops $R$ and $B$.} } \label{fig1.fig}
 \end{figure}

Consider the Verhulst model \citep{murray2002mathematical,kunsch2006verhulst}, which describes the growth of a population $x$ in an environment with a  carrying capacity $M$, and  per capita rate of growth in the absence of capacity effects $r$.  As a differential equation model it is often represented by the logistic equation
 \begin{equation}
 \dot{x} = r x(1-x/M) \label{verhulst0.eq}
 \end{equation} 
which can be solved in closed form. By contrast, system dynamics represents this model as a \emph{set} of equations, for example: 
\begin{equation}
\left.
\begin{array} {rcll}
\dot{x} &=& G \\
 G&=&gx&\mbox{rate of growth} \\
 g&=&rf&\mbox{per capita rate of growth} \\
f&=&1-x/M&\mbox{fractional shortfall of population from carrying capacity}\\
\end{array}
\right\} \label{verhulstsystem.eq}
\end{equation} 
which represent the assumptions of the model. In system dynamics, the dynamic variable $x$ is called a \emph{stock}, emphasising that it is an \emph{accumulation} of material, in this case people, resulting from the integration of the differential equation in (\ref{verhulstsystem.eq})\footnote{The terms ``dynamic variable'' and ``stock'' will be used interchangeably in this paper.}.  This set of  equations, (\ref{verhulstsystem.eq}), is expressed in a \emph{stock/flow} diagram, figure \ref{fig1.fig}, that highlights the model's causal structure.  $G$, the rate of change of $x$, is referred to as a flow. The variables $g$ and $f$ are called auxiliaries. They depend algebraically on other variables.  The single arrows are called connectors and indicate a causal connection, described by an algebraic formula, between source and target variables. The signs on the connectors indicate whether the target variable changes the same way as its cause (+), or the opposite way (-). By contrast, the sign on the flow (+) indicates accumulation.   The model diagram makes explicit the two feedback mechanisms contained in the underlying four equations  (\ref{verhulstsystem.eq}): a reinforcing loop $R$ of positive polarity, representing the growth process, and a balancing loop $B$ of negative polarity representing saturation effects as the population approaches capacity\footnote{The polarity of a feedback loop is positive if an increase in the stock value leads to an increase in the net flow rate to the stock. Likewise, negative loop polarity is where a stock increase results in reduced net flow rate. If there are many stocks in a loop, then such changes are delayed due to transients \citep{richardson1995loop}.  }.  The loop polarities are determined by the product of the connector and flow polarities. The four equations, and the network connections they embody, represent the ``physics'' of the model, i.e. the causal  relationships between variables. 

Equation (\ref{verhulst0.eq}) is not the only form of the Verhulst model. For example, the model can also be presented as $\dot{x} = r x-rx^2/M $. However, unlike (\ref{verhulst0.eq}), this form of the equation has lost the connection with the model assumptions contained in the  four equations (\ref{verhulstsystem.eq}); thus it fails to capture the model's  causal network, figure \ref{fig1.fig}.  Although such re-arrangements of equations are useful for model analysis, the resulting loss of the original  network information makes it hard to identify the causal links and feedback loops -- the ``physics'' of the system dynamics model, figure \ref{fig1.fig}. Indeed, although differential equation modelling is considered a branch of sociophysics,  its connection with physics is less clear compared with modelling that uses,  for example, statistical mechanics, or thermodynamics. 

Recently, a technique to measure the relative effects of causal connections in system dynamics models has been introduced by \citet{hayward2014model}, called the \emph{loop impact method}. This method allows for the causal pathways, such as those represented by loops $R$ and $B$ in figure \ref{fig1.fig}, to be interpreted as forces in the Newtonian sense, with the behaviour of the dynamic variables being explained by the balance of forces acting on each one. \citet{hayward2017newton} extended the work, showing that the ``physics'' of a model's causal connections can be interpreted in Newtonian mechanical terms with the differential equations being re-expressed in a form similar to Newton's laws of motion. This understanding is called the \emph{Newtonian Interpretive Framework}.  Key to this framework is the concept of the \emph{impact} of a force, measured by the ratio of the acceleration produced by the force on a state variable $x$ to the rate of change of $x$, that is $\dd x/\dd t$, the analogy of velocity. The impact of a force highlights the connection between force and the feedback resulting from causal loops \citep{hayward2014model}, and describes the curvature in the time graph of state variables so that exponential behaviour has constant impact \citep{hayward2017newton}. To date, this method and framework have only been used within the system dynamics community.  The purpose of this paper is to apply the loop impact method to population modelling using differential equations, thus highlighting the sociophysics in the models through the Newtonian Framework, especially the concept of force.

One of the strengths of Forrester's system dynamics is its ability to provide explanations of model behaviour in terms of the causal structure of the model \citep{sterman2000business}. In particular, a number of tools have been developed to examine behaviour in terms of feedback loops, \citep{duggan2013methods,kampmann2008structural}. \possessivecite{hayward2014model} loop impact method is a development of one of these tools, the pathway participation metric \citep{mojtahedzadeh2004using}, with a more comprehensive method for comparing causal effects. However, the loop impact method also interprets the causal connections as forces which can be computed analytically, without the reliance on numerical simulation \citep{hayward2017newton}. As such, the method is easily translated into any modelling application that uses systems of ordinary differential equations, such as population, biological or social modelling, allowing for Newtonian interpretations of model behaviour.

The concepts in the loop impact method are evident in some other methodologies. For example, feedback is used in biological systems \citep{thomas1990biological,gouze1998positive,boker2001differential,cinquin2002roles,deangelis2012positive,mooij2013ordinary}, where the combination of feedback loops is related to the number, and nature, of equilibrium states. However, the methods used do not quantify the effect of individual feedback connections on variables over time, nor do they interpret the results in Newtonian mechanical terms. Newtonian concepts have been quantified in a social context by a number of authors, e.g. social impact theory \citep{latane1981psychology}, the social force model of  pedestrian dynamics \citep{helbing1995social}, and the social force of Montroll \citep{montroll1978social,ausloos2013another}. However, the quantifications used do not follow the classical Newtonian construction. By contrast, the loop impact method defines the force between populations using a Newtonian formulation such that the force due to a feedback loop causes variables to accelerate or decelerate. Thus there is a direct connection between the model feedback structure and the behaviour of dynamic variables. This approach could be described as \emph{sociomechanics}.

The paper is structured as follows: Section 2 describes the loop impact method and the Newtonian Interpretive Framework, defining the concept of force in systems of differential equations using the causal structure of system dynamics. In subsequent sections the  framework is applied to the Spruce Budworm model, a first-order example, and a predator-prey model, a second-order model,  with discussion of  the force dominance using the loop impact method.

\section{Force, Impact and Newtonian Framework}
\subsection{Force} \label{force.sec}
In Newtonian mechanics a force $F$ is a cause that produces acceleration in a body of mass $m$, $\ddot{x} = F/m$, where $x$ is the location of the body. In this expression of Newton's second law of motion, the direction of causality is right to left. The higher the mass,  the less effect the force has on the motion of the body. If there is no force then $\ddot{x}=0$ and the body is either at rest or undergoing uniform motion, Newton's first law of motion.  Uniform motion is typically represented as a straight line graph of $x$ against time, whereas if there is a force then the graph has curvature. The greater the force, the larger the curvature, that is, the more the trajectory deviates from a straight line. 

Likewise, the graphical solution of a set of non-linear first-order ODEs (\ref{ode.eq}) will exhibit curvature for $x_i$ against time as, in general, its second derivatives are non-zero.  Differentiating (\ref{ode.eq}) gives:
 \begin{equation}
 \ddot{x_i} = \sum_j \frac{\partial f_i }{ \partial x_j }\dot{x}_j \label{jacobi.eq}
 \end{equation}
  showing the dependency of the curvature on the Jacobian $\partial f_i / \partial x_j$. By analogy with Newton's second law of motion, the right hand side of (\ref{jacobi.eq})  suggests that the elements of the Jacobian play the role of forces, causing acceleration in the dynamic variables $x_i$. These forces originate in the variables themselves. However equation (\ref{jacobi.eq}) lacks sufficient detail for identification of the individual forces, as the form of (\ref{ode.eq}) does not distinguish multiple causal pathways between specific variables. For example, the Verhulst model, figure \ref{fig1.fig}, has two causal pathways associated with loops $R$ and $B$. 
  While the differential equation of the model, $\dot{x} = r x(1-x/M)$, embodies clearly the assumptions of the model, it is not adequate to distinguish the two causal pathways. Thus the two  forces cannot be distinguished in its Jacobian $\partial f/\partial x = r(1-2x/M)$.  Therefore, in this differential equation form, it is impossible to determine the contribution of each force, i.e. feedback loop, to the curvature in the graph of $x$ against time.

To proceed, the ODEs (\ref{ode.eq}) need to be expressed in network form, preserving the causal topology of the system dynamics model, e.g. figure  \ref{fig1.fig}, which encapsulates the model assumptions.  Following \citet{hayward2017newton},
 let there be $\pi_{ij}$ causal pathways from  stock $x_i$ to  stock $x_j$.  Let $\mu_{ij}=1,  \ldots, \pi_{ij}$ index the pathways between a given pair of stocks. Thus, all the pathways in the model can be assigned a label $a_{ij\,\mu_{ij}}$, which can be abbreviated to  $a_{ij \,\mu}$ without confusion. Therefore, $a_{ij}$ is a matrix of vectors, of possibly differing dimension $\pi_{ij}$. The vector index $\mu_{ij}$ will be used to distinguish pathways between the same pair of stocks, whereas the matrix $a_{ij}$ that contains the vectors will distinguish pathways between different pairs of stocks.

Thus, a general nth order system dynamics model is a given by 
\begin{equation}
\frac{\dd x_i}{\dd t
} =    f_i( x_{j\underline{a_{ji\,\mu}}}) \;\;\;\;  i,j = 1, \ldots, n; \;\;  \mu_{ji} = 1, \ldots, \pi_{ji} \label{sd.eq}
 \end{equation}
  where $ x_{j\underline{a_{ji\,\mu}}}$ is the variable $x_j$ along pathway $a_{ji\,\mu} \equiv a_{ji\,\mu_{ji}}$ connected to $x_i$. Thus the pathway name has been used as an index on the variable name in order to distinguish the source of a particular causal connection, and thereby label a specific force. This index is underlined to distinguish it from indices used to label dynamic variables (stocks). The equations (\ref{sd.eq}) are referred to as \emph{causally connected differential equations} and represent symbolically a system dynamics model \citep{hayward2017newton}.  The terms $f_i$ are the net flows on each stock.
  
  For example, the Verhulst model, figure \ref{fig1.fig}, can be expressed as the causally connected ODE:
  \begin{equation}
 \frac{\dd x}{\dd t
} = rx_{\underline{G}}\left(1 - \frac{x_{\underline{fgG}}}{M}\right) \label{verhulst.eq}
  \end{equation} 
where the two causal pathways from stock $x$ to its own flow have been labelled with their auxiliary variables, $a_{11\,1} = G$ and $a_{11 \,2}= fgG$. Such pathway labels are unique and can be deduced from the model equations (\ref{verhulstsystem.eq}), see \ref{appendix1}\footnote{It is sometimes possible to reduce the number of auxiliary variables in the pathway label or even substitute the loop name, see  \ref{appendix1} and \citet{hayward2017newton}. To avoid confusion, this paper uses all auxiliary variables names.}.  Thus the two pathways are distinguished in (\ref{verhulst.eq}), enabling the two forces to be quantified. Examples of the notation in models with more than one stock are given in sections \ref{forcehigher} and \ref{predprey}.

The forces along each pathway in the general model are determined by differentiating  (\ref{sd.eq}) with respect to time and treating each pathway labelled variable $ x_{j\underline{a_{ji\,\mu}}}$ as an independent variable. Thus, differentiating along each causal pathway: 
 \begin{equation}
  \ddot{x}_i = \sum_{j=1}^n \frac{\partial f_i}{\partial x_j}  \dot{x}_j = \sum_{j=1}^n \; \sum_{\mu_{ji}=1}^{\pi_{ji}}  \left. \frac{\partial f_i}{\partial x_j} \right \|_{\underline{a_{ji\,\mu}}}  \dot{x}_j  \label{sddiff.eq}
  \end{equation} 
  where the pathway derivative is defined by:
 \begin{equation}
   \left. \frac{\partial f_i}{\partial x_j} \right \|_{\underline{a_{ji\,\mu}}} \triangleq \frac{\partial f \;\;\;\;}{\partial x_{j\,\underline{a_{ji\,\mu}}}}  \label{pathway.eq}
   \end{equation}
  which is the derivative along one pathway $a_{ji\,\mu}$.

 \subsection{Impact} 
 Rather than deal with the force itself, \citet{hayward2014model} showed that the  \emph{impact} of the force, defined as the ratio of the acceleration of a variable to its rate of change,  was a more appropriate measure of force as it  preserves the connection between force and the polarity of feedback loops. Expressing (\ref{sddiff.eq}) in impact form gives:
  \begin{equation}
  \frac{\ddot{x}_i}{\dot{x}_i} =\sum_{j=1}^n \; \sum_{\mu_{ji}=1}^{\pi_{ji}}  \left. \frac{\partial f_i}{\partial x_j} \right \|_{\underline{a_{ji\,\mu}}}  \frac{\dot{x}_j}{\dot{x}_i}  \label{impact1.eq}
  \end{equation} 
 Thus, the impact, denoted $\Ip$, of the force of $x_j$ on $x_i$ along a specific pathway $a_{ji\,\mu}$ is:
  \begin{equation}
 \Ip_{\underline{x_j a_{ji\,\mu}x_i}} \triangleq \left. \frac{\partial f_i}{\partial x_j} \right \|_{\underline{a_{ji\,\mu}}} \frac{\dot{x}_j}{\dot{x}_i} \label{impact2.eq}
  \end{equation} 
  where the underlined subscript on $\Ip$ indicates the source $x_j$, pathway $a_{ji\,\mu}$ and target $x_i$ of the force\footnote{$\Ip_{\underline{x_j a_{ji\,\mu}x_i}} $ is read: the impact of the force of $x_j$ on $x_i$ via pathway $a_{ji\,\mu}$.}. 
  
\begin{figure}[!ht]
          \begin{center}
   \includegraphics[height=5.5cm] {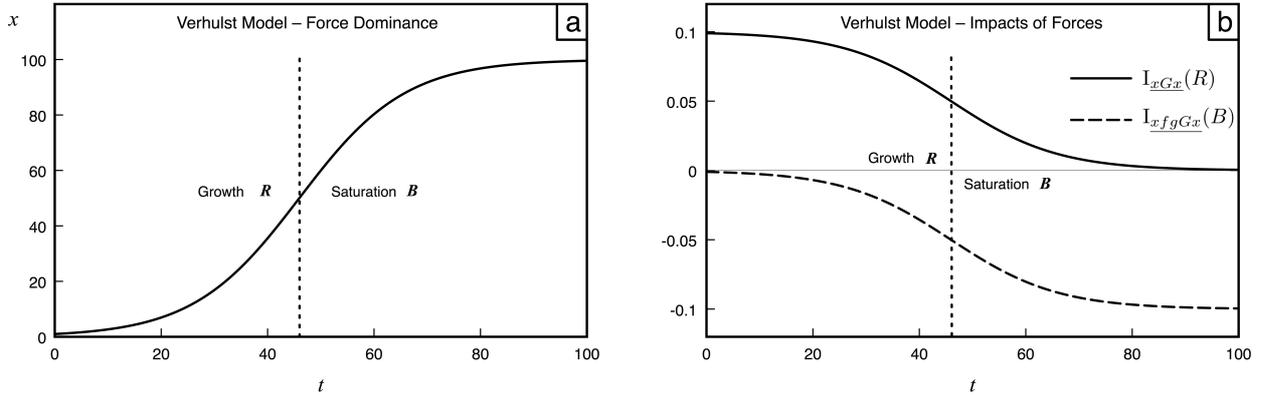}    
       \end{center}
    \vspace{-15 pt}
    \caption{\small{(a) Logistic solution to Verhulst model for population $x$ and regions of force dominance associated with loops $R$, $B$; $r=0.1$, $M=100$ and $x_0=1$. (b) Impacts of forces: $\Ip_{\underline{xGx}}(R),\Ip_{\underline{xfgGx}}(B)$. } }  \label{fig2.fig}
 \end{figure}

For example, the impacts of the two forces on $x$ in the Verhulst model (\ref{verhulst.eq}) are obtained using pathway differentiation on $x_{\underline{G}}$  and $x_{\underline{fgG}}$:
  \begin{equation}
 \Ip_{\underline{xGx}} (R) = r\left(1 - \frac{x}{M}\right), \,\,\,\,\,\, \,\,  \Ip_{\underline{xfgGx}} (B)  = -r\frac{x}{M}\label{verhurlstimpact.eq}
  \end{equation} 
 where the pathway labels on $x$ have now been dropped. For clarity, the loop names associated with the two forces have been given in brackets\footnote{$\Ip_{\underline{xfgGx}} (B)$ is read: the impact of the force of $x$ on itself via pathway $fgG$ associated with feedback loop $B$.}. The two forces have equal effect  on the curvature of $x$ when $\left|\Ip_{\underline{xGx}}\right(R)|=\left|\Ip_{\underline{xfgGx}}(B)\right|$, which occurs at $x=M/2$, the inflexion point,  figure \ref{fig2.fig}a.  For $x<M/2$ the impact of the force via $R$ is the greater, resulting in accelerating growth. For $M/2<x\le M$ the impact of the force via $B$ is greater, resulting in deceleration to the carrying capacity.  Thus, the logistic behaviour of the Verhulst model is explained by a change in force dominance, also called loop dominance in system dynamics \citep{kunsch2006verhulst,richardson1995loop,senge2006fifth}. 
 
 One advantage of using the impact ratio to measure force is that the sign of the impact preserves the polarity of the corresponding feedback loop. In this example, as both loops are of first order, the impact is the same as the loop gain \citep{hayward2014model}. Thus, in the Verhulst model for $x>0$,   $\Ip_{\underline{xGx}}(R)>0$,  reflecting the positive polarity of a reinforcing loop, whereas $\Ip_{\underline{xfgGx}}(B)<0$, the negative polarity of a balancing loop, figure \ref{fig2.fig}b.

 A positive impact force always imparts acceleration and a negative impact force always imparts deceleration, regardless of whether $x$ is increasing or decreasing.  
 Consider $\dot{x}=ax-h$ where $a>0$ is a growth rate, and $h$ is constant harvesting. The only force is due to growth with impact $\Ip_{\underline{xx}} = a$, which is always positive,  a reinforcing effect. This force causes acceleration if the initial rate of change is positive, $\dot{x}_0>0$, i.e. $ax_0>h$. If, however, the  initial rate of change is negative, $ax_0<h$, then  the force with positive impact causes $x$ to have accelerating decline, because the declining population becomes less able to replenish numbers lost by the constant harvesting. In this sense a ``growth'' force can be said to cause accelerating decline. Note that the acceleration is negative, but impact remains positive as it is a ratio with rate of change. Thus the polarity is always associated with either acceleration (positive) or retardation (negative). 

A second advantage of using impact to measure force is the association between constant impact and linear processes. Consider the exponential model, $\dot{x}=ax$, which has a single linear reinforcing loop (implicit in the equation).  The impact of the single force due to the reinforcing loop is constant, $\Ip_{\underline{xx}} = a$. Thus impact, the ratio of a stock's acceleration to its rate of change, is a measure of curvature such that an exponential curve has constant impact. Thus, linear first-order models have constant impacts. A non-linear first-order model, such as the Verhulst model, will have variable impacts, figure \ref{fig2.fig}. The variability of the impact indicates the extent to which a stock's behaviour deviates from exponential. 

Impact is a fractional measure of curvature with units of inverse time (\ref{impact1.eq}) and  thus its units are  independent of those of either target or source variables. This feature enables forces, to and from different variables, to be directly compared regardless of their units,  a  third advantage of using impact as a measure of force. Thus, in a model of two or more variables, the balance of forces contributing to the curvature of a variable can still be compared with behaviour explained by the dominant forces, the loop impact method of \citet{hayward2014model}. 

\subsection{Newtonian Interpretive Framework}\label{NIF.sec}
In the  Newtonian Interpretive Framework of \citet{hayward2017newton}, the forces on the variables,  indicated by the system dynamics model, are interpreted by analogy with mechanics.  For example, consider  a linear first-order balancing loop, $\dot{x}=-ax$, where $a>0$. The single force in this system is frictional: $\ddot{x} +a \dot{x}=0$, where $a$ controls the amount of dissipation of the material in stock $x$. By contrast, a first order reinforcing loop, $\dot{x}=ax$, where $a>0$, is a self-generating force. Thus, the Verhulst model could be interpreted as the balance of two forces with non-constant impact (\ref{verhurlstimpact.eq}): a self-generating mechanism that decreases with increasing $x$, being opposed by friction that increases as $x$ approaches capacity. 

\begin{figure}[!ht]
          \begin{center}
   \includegraphics[height=3.2cm] {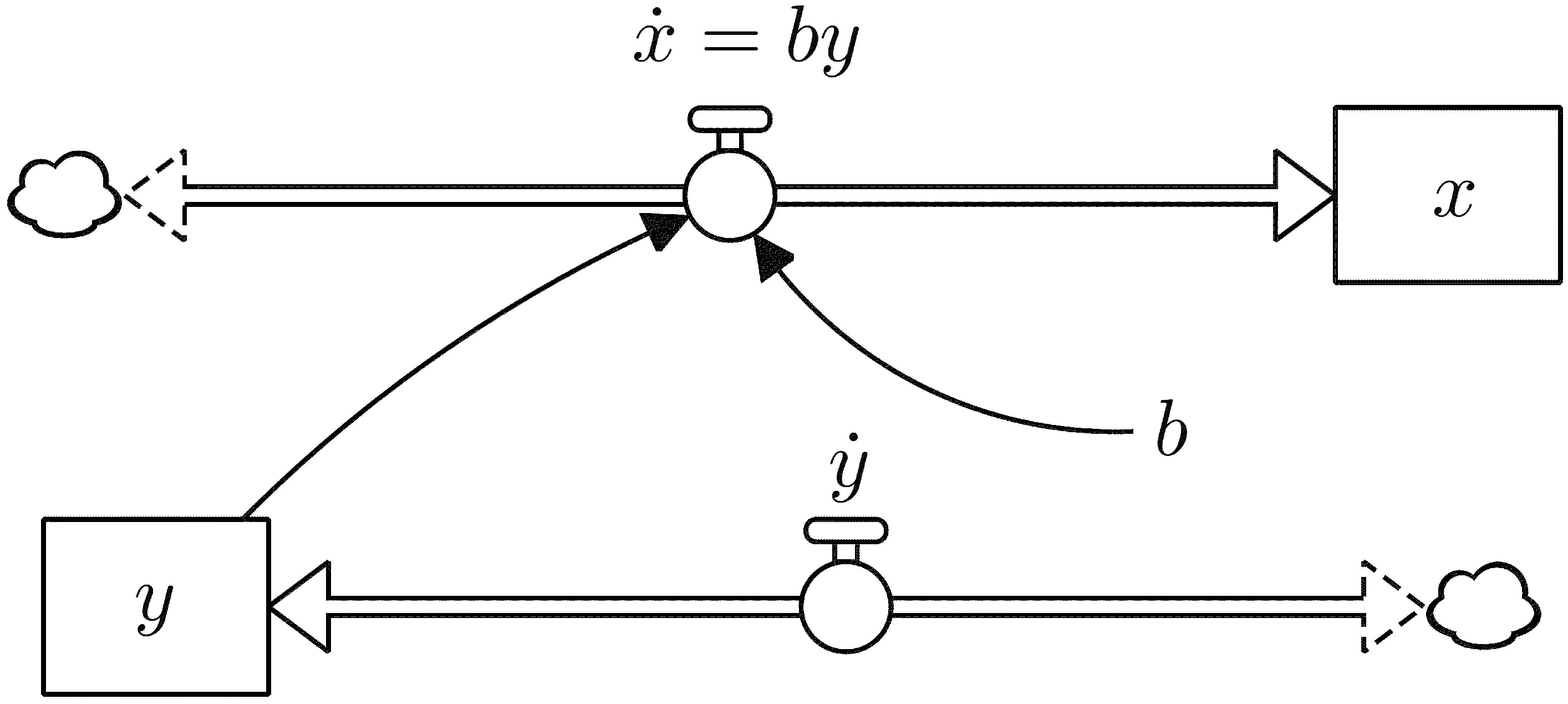}
       \end{center}
    \vspace{-15 pt}
    \caption{\small{Variable $x$ influenced by variable $y$. $\dot{y}$ quantifies the force of $y$ on $x$. $x$ has momentum $y$ and mass $b^{-1}$.} } \label{fig3.fig}
 \end{figure}

In higher order models there will be forces between different variables.  Consider  $\dot{x} = by(t)$, where $x$ and $y$ are both stocks, figure \ref{fig3.fig}.  The variable $y$ exerts a force on $x$ as $\ddot{x} = b\dot{y}$. Thus changes in $y$  are associated with acceleration in $x$. Thus, following the Newtonian Interpretive Framework, $\dot{y}$, the net flow on $y$, measures the  force of $y$ on $x$ \citep{hayward2017newton}. The coupling parameter $b$ represents the inverse of the ``mass'' of $x$ with respect to $y$, $m\triangleq b^{-1}$. Thus, the acceleration of  $x$ is given by the equivalent of Newton's second law: $\ddot{x} = (1/m) F = b \dot{y}$, where mass converts force $F=\dot{y}$ into acceleration. It follows that if $b$ is large, then the mass of $x$ is small, and thus only small changes in $y$ are needed to accelerate $x$. However, if $b$ is small, $x$ is heavier with respect to $y$ and has more inertial resistance to change in its motion. Given this interpretation of the coupling parameter $b$ as the inverse of mass, then the initial value of variable $y$ represents the initial ``momentum'' of variable $x$, as $\dot{x} = (1/m) y$.    

An example will help illustrate the Newtonian framework. Consider a constant force $F_c$ acting on $x$, i.e. $y=y_0 + F_c t$, where $y_0$ is the initial momentum of $x$. Further, let this force be negative, $F_c = -k$, acting to reduce the rate of change of $x$ to zero.  Thus $\dot{y}=-k$, which represents the force of $y$ on $x$. It follows that the rate of change of $x$ is $\dot{x} = by_0 - bkt$, giving $x= by_0t - \frac{1}{2}bkt^2$, assuming $x_0=0$. The variable $x$ is momentarily stationary, $\dot{x}=0$, at $t=y_0/k$, and thus for a fixed force $k$, the higher the initial value of $y$, the longer it takes to achieve $\dot{x}=0$. The Newtonian framework interprets this longer time to rest as being due to the high momentum $y$ of the stock $x$. The same time to rest can be achieved if a higher force is used to overcome the higher momentum.
At rest, $x=by_0^2/(2k)=y_0^2/(2mk)$. Thus, the higher the mass of variable $x$, i.e. the lower the coupling $b$ from $y$, the smaller the value achieved by $x$. A high mass variable does not change as much as a low mass variable because it has more inertial resistance.  Thus, the variable $x$ in figure \ref{fig3.fig} has momentum and mass with regard to $y$'s influence~\citep{hayward2017newton}.

 From (\ref{impact2.eq}), the impact of the force of $y$ on $x$ is:
 \[
 \Ip_{\underline{yx}} = \frac{\partial\dot{x}}{\partial y} \frac{\dot{y}}{\dot{x}} = -\frac{k}{y_0-kt}
 \]
 giving negative impact for $t<y_0/k$. Because of its ratio nature, impact tends to infinity as $x$ gets closer to equilibrium, $\dot{x}=0$. Once $x$ starts accelerating again for $t>y_0/k$,  the impact of $y$ on $x$ is positive as $x$ is now decreasing in value with a negative force. Change of impact polarity is an important interpretive tool when explaining the behaviour of systems with higher order feedback loops, i.e. those containing  two or more variables.

\subsection{Force and Higher Order Feedback} \label{forcehigher}
\begin{figure}[!ht]
          \begin{center}
   \includegraphics[height=3.5cm] {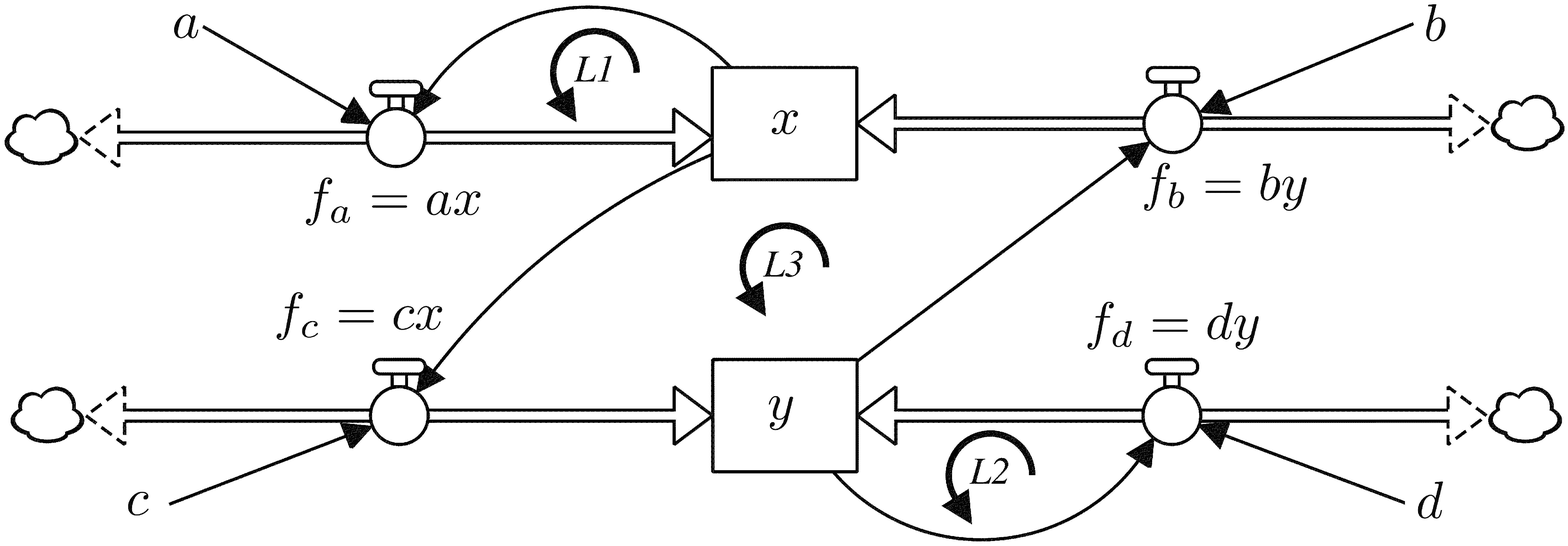}
       \end{center}
    \vspace{-15 pt}
    \caption{\small{General second-order linear system, with feedback loops $L_1$, $L_2$, and $L_3$. $x$, $y$, $a$, $b$, $c$ and $d$ may take any real value. }} \label{fig4.fig}
 \end{figure}

To illustrate  the concept of force for models with two or more variables, consider the general second-order linear system: $\dot{x}=ax+by$, $\dot{y}=cx+dy$, where $a,b,c,d$ are constants. Variables and constants may be positive or negative.  The system has one equilibrium point $(0,0)$ whose stability is determined by the constants. These equations can be expressed in system dynamics form, figure \ref{fig4.fig} and equations (\ref{lin1.eq}--\ref{lin2.eq}), where the pathways between the variables are labelled by the flow names between source and target stocks. In terms of the general labelling (\ref{sd.eq}), with $x$ and $y$ labelled 1 and 2 respectively: $a_{11 \, 1} = f_a$, $a_{21 \, 1} = f_b$, $a_{12 \, 1} = f_c$, $a_{22 \, 1} = f_d$. In these equations there is only one causal link between each pair of stocks, thus $\pi_{ij} = 1, \forall_{ij} $. 
\begin{eqnarray}
\frac{\dd x}{\dd t} &=& ax_{\underline{f_a}}+by_{\underline{f_b}} \label{lin1.eq} \\
\frac{\dd y}{\dd t} &=&cx_{\underline{f_c}}+dy_{\underline{f_d}} \label{lin2.eq}
\end{eqnarray}

From the perspective of the Newtonian framework, each variable is subject to two forces, one from itself, and one from the other variable. The self-forces are associated with the two first-order feedback loops $L_1,L_2$, with gains $G_1 = a, G_2 = d$, \citep{hayward2017newton,kampmann2012feedback,richardson1995loop}.  The two forces from $x$ to $y$ and vice versa form a second-order feedback loop $L_3$, whose gain is $G_3 = bc$. Thus, the polarities of the loops $L_1$ and $L_2$ are determined by the signs of $a$ and $d$, whereas the polarity of $L_3$ is determined by the sign of the product of $b$ and $c$.

The gains of the loops determine the eigenvalues of the system:
\begin{equation}
\lambda_{\pm} = \tfrac{1}{2}\left[ G_1+G_2 \pm\sqrt{(G_1-G_2)^2+4G_3} \right] \label{ev.eq}
\end{equation}
and, as such, determine the growth/decay constants and oscillation frequencies \citep{hayward2017newton,kampmann2012feedback,richardson1995loop}.  For example, if $a=d=0$ and $b$ and $c$ have opposite signs, the system has a single second-order balancing loop and oscillates with frequency $\sqrt{|G_3|}$. If $a$ and $d$ are non-zero, then the first-order loops are active, and the system  either grows or decays exponentially  to equilibrium with exponent $(G_1+G_2)/2$.

The well-known stability criteria of a second-order linear system can be expressed in loop gains. The point $(0,0)$ is stable if and only if $G_1+G_2< 0$ and $G_1G_2>G_3$ \citep{hayward2017newton,drazin1992nonlinear}. Thus, to ensure stability, at least one of the first-order loops must be balancing. As already shown, such a loop represents a frictional force, and thus, in the Newtonian framework, the system must have sufficient friction to counteract the forces due to the remaining loops. The system  (\ref{lin1.eq}--\ref{lin2.eq}), figure \ref{fig4.fig},  can only be stable  if either $L_3$ is reinforcing and both $L_1$ and $L_2$ are balancing; or $L_3$ is balancing with at least one of the first-order loops also balancing. Regardless of stability, the system oscillates if $(G_1-G_2)^2+4G_3 <0$. Thus the minimum condition for an oscillating system is that the second-order loop is balancing, $G_3<0$.

Although the gains determine the stability criteria, they alone do not describe the full behaviour of $x$ and $y$, which also depend on their initial values.The shape of these variables' trajectories over time can be explained by examining the balance of forces on each variable, where the forces are measured by the loop impacts.  Using (\ref{impact2.eq}) and (\ref{pathway.eq}), the four impacts in  (\ref{lin1.eq}--\ref{lin2.eq}), figure \ref{fig4.fig} are:
\begin{eqnarray}
\Ip_{\underline{xf_ax}}(L_1) &=& \left. \frac{\partial \dot{x}}{\partial x} \right \|_{\underline{f_a}}  =  a \label{linearimpact.1} \\
\Ip_{\underline{yf_dy}}(L_2) &=&  \left. \frac{\partial \dot{y}}{\partial y} \right \|_{\underline{f_d}} =d \label{linearimpact.2} \\
\Ip_{\underline{yf_bx}} (L_3)&=&  \left. \frac{\partial \dot{x}}{\partial y} \right \|_{\underline{f_b}} \frac{\dot{y}}{\dot{x}}= \frac{b(cx+dy)}{ax+by} \label{linearimpact.3} \\
\Ip_{\underline{xf_cy}} (L_3)&=&  \left. \frac{\partial \dot{y}}{\partial x} \right \|_{\underline{f_c}}   \frac{\dot{x}}{\dot{y}}= \frac{c(ax+by)}{cx+dy} \label{linearimpact.4} 
\end{eqnarray} 
For the two first-order loops, $L_1$, $L_2$, the impacts (\ref{linearimpact.1}--\ref{linearimpact.2}) are the same as the loop gains $G_1,G_2$. The product of the second-order impacts, (\ref{linearimpact.3}--\ref{linearimpact.4}),   is the loop gain of $L_3$:  $\Ip_{\underline{yf_bx}}(L_3)\Ip_{\underline{xf_cy}}(L_3)=bc=G_3$. This result is a particular case of the loop impact theorem, which states that, for a system of any order, complexity, or non-linearity, the product of all impacts in a loop equals the loop gain \citep{hayward2014model}. Thus, the loop gain is ``shared'' between the two forces of $x$ on $y$ and vice versa. Although the gain retains its polarity, reinforcing or balancing, the signs of the impacts may change, allowing the second-order loop to accelerate a stock in some periods and decelerate it in others. In each case, the force on the other stock in the loop will  have the appropriate polarity to preserve the sign of the loop gain. 

\begin{figure}[!ht]
          \begin{center}
   \includegraphics[height=11.3cm] {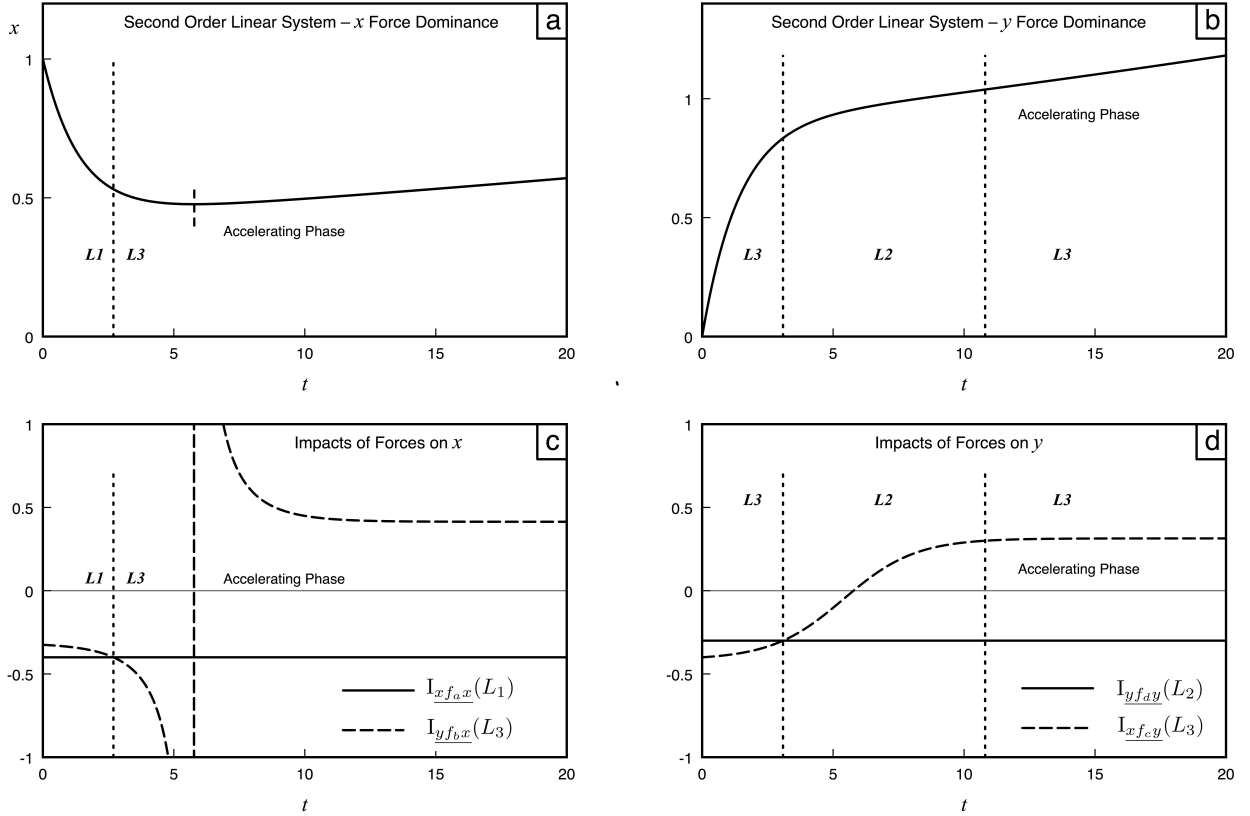}
       \end{center}
    \vspace{-15 pt}
    \caption{\small{Simulation of second-order linear system with $a=-0.4$, $b=0.2$, $c=0.65$,  $d=-0.3$, $x_0=1$ and $y_0=0$. The system is unstable with $G_1G_2<G_3$. (a) $x$ force dominance. (b) $y$ force dominance. (c) Impacts of the forces on $x$, $\Ip_{\underline{xf_ax}}(L_1)$, $\Ip_{\underline{yf_bx}}(L_3)$. (d) Impacts of the forces on $y$, $\Ip_{\underline{yf_dy}}(L_2)$, $\Ip_{\underline{xf_cy}}(L_3)$. }} \label{fig5.fig}
 \end{figure}

Consider the system (\ref{lin1.eq}--\ref{lin2.eq}) where $a=-0.4$, $b=0.2$, $c=0.65$, $d=-0.3$, and $(x_0,y_0) = (1,0)$. Here there are two first-order balancing loops, i.e. frictional forces, opposing the self-generating forces associated with the second-order loop. The system is a saddle, and therefore unstable with no oscillations. The trajectories for $x$ and $y$ and their impacts (\ref{linearimpact.1}--\ref{linearimpact.4}) are given in figure \ref{fig5.fig}.

Initially, $x$ is declining and slowing down with its frictional force $L_1$ dominating, figure \ref{fig5.fig}a. By contrast, $y$ is increasing and also slowing down with the second-order reinforcing loop $L_3$ dominant,  figure \ref{fig5.fig}b. In this initial period, the reinforcing loop has negative polarity on both variables and is thus causing deceleration,  figures \ref{fig5.fig}c--\ref{fig5.fig}d. From $t=2.8$,  $L_3$ begins to dominate on $x$, and  at $t=3.1$, $L_2$ dominates on $y$, both still slowing

The impacts of loop $L_3 $ change polarity from negative to positive at $t=5.8$, causing $x$ to be momentarily stationary and then start increasing, figure \ref{fig5.fig}a. At this point, the impact of $L_3$ on $y$ is zero, figure \ref{fig5.fig}d; thus all $L_3$'s force is transferred to $x$ with infinite impact, figure \ref{fig5.fig}c. This change of link polarity accounts for the dominance of $L_3$ on $x$, and $x$'s change of direction. As the force of $L_3$ on $y$ grows, it eventually dominates over the frictional force, causing $y$ to accelerate with the inflexion point at $t=10.8$. 

Thus the shapes of the curves of the variables against time are determined by the balance of forces associated with the causal connections between the variables. Both $\Ip_{\underline{yf_bx}}(L_3)$ and $\Ip_{\underline{xf_cy}}(L_3)$ tend to a limit due to the ratio definition of impact. For the parameters in figure \ref{fig5.fig}, these values remain numerically larger than those of  the first-order loops; thus these frictional forces are unable to overcome the force of the second-order loop. In general, for a system with a second-order reinforcing loop and two first-order balancing loops, it can be shown that the condition for the frictional forces to overcome the reinforcing loop is equivalent to the stability criterion. For example, when $t\rightarrow \infty$, $x$ and $y$ tend to exponential behaviour in the dominant eigenvalue, $\lambda_+$, thus in the limit,  $\Ip_{\underline{yf_bx}}(L_3)=c\lambda_+/(c+d(\lambda_+-a)/b)$. It follows that,  if $\Ip_{\underline{yf_bx}}(L_3)+\Ip_{\underline{xf_ax}}(L_1)<0$ as $t\rightarrow \infty$, the condition for friction to ultimately dominate, then it must be the case that $G_1G_2>G_3$, the condition for stability.  Therefore, bifurcation is explained by the balance of forces associated with changes in loop impact.

According to the Newtonian Interpretative Framework, section \ref{NIF.sec}, the coupling constants between variables can be interpreted as the inverse of masses, see figure \ref{fig3.fig}. Thus, for the two variable linear system, $b^{-1}$ is the mass of $x$ with respect to $y$, and $c^{-1}$ the mass of $y$ with respect to $x$, (\ref{lin1.eq}--\ref{lin2.eq}),  figure \ref{fig4.fig}. Thus, the gain of the second-order loop $G_3=bc$ is taken to represent the inverse of the mass of that loop. In this sense, the magnitude of the inverse of the loop gain $|G_3^{-1}|$ represents the inertia of the second-order feedback loop, and thus, in this case,  the system.  For example, in the simulation in figure \ref{fig5.fig}, where $L_3$ is reinforcing, then the system has insufficient inertia for its friction to overcome the driving force of the reinforcing feedback loop, $G_1G_2<G_3$. Expressed more informally: the system is too ``light''  to control the reinforcing loop due to the high gain $G_3$.  In a ``heavier system'', smaller $G_3$, the force of $L_3$ is less effective on the higher mass and stability is achieved.     By contrast, if  $L_3$ is balancing and the system is oscillating, then a higher mass system will result in a smaller frequency of oscillations,  (\ref{ev.eq}), and  is thus more sluggish. Therefore mass is a useful interpretative concept for feedback loops with two or more variables.

This section has described how population models based on ordinary differential equations can be interpreted in Newtonian mechanical terms, with variable behaviour explained by comparing the forces between populations. This interpretive framework expresses the models using the causal network form of system dynamics where feedback between variables is made explicit. It is this feedback that is interpreted as force, utilising the loop impact method. This framework is next applied to two standard population models, to demonstrate the benefits of employing the force concept in addition to equilibrium analysis.


 \section{Spruce Budworm Model} \label{spruce.sec}
 \subsection{Model and Equilibrium Analysis}
\citet{ludwig1978qualitative} proposed a model of the growth of a spruce budworm population,  density $x$, who feed on the leaves of balsam fir trees and are themselves subject to predation by birds. In this simplified model, the  carrying capacity $M$ is assumed constant because budworm densities change much faster than leaf area. Predation is modelled by a Holling type III function due to the birds seeking food elsewhere when budworm numbers are low \citep{murray2002mathematical,freedman1980deterministic,ludwig1978qualitative,holling1959components,holling1966functional,strogatz2018nonlinear}.  Thus, the model is represented by a first-order differential equation, the Verhulst model with predation (\ref{budworm.eq}):
\begin{equation}
\frac{\dd x}{\dd t}=  r x \left( 1-\frac{x}{M}\right) - \beta \frac{x^2}{\alpha^2 + x^2} \label{budworm.eq}
\end{equation}
where $r$ is the per capita growth rate in the absence of capacity effects, $\beta$ is the maximum predation rate, and $\alpha$ controls the scale of budworm densities at which saturation begins to take place.  For analysis, the number of parameters can be reduced to two by making $x$ and $t$ non-dimensional \citep{murray2002mathematical,strogatz2018nonlinear}. Thus, without loss of generality,  $\alpha = \beta=1$.

The equilibrium analysis of the budworm differential equation (\ref{budworm.eq}) is well known  and  exhibits different patterns of growth  associated with the number of equilibrium points \citep{murray2002mathematical,strogatz2018nonlinear}. Setting $\dot{x}=0$ in  (\ref{budworm.eq}) gives the extinction equilibrium point $x=0$, which is always unstable for $r>0$, and the non-zero equilibrium points determined by the solution of (\ref{budeqp.eq}):
\begin{equation}
r  \left( 1-\frac{x}{M}\right) = \frac{x}{1 + x^2} \label{budeqp.eq}
\end{equation}

\begin{figure}[!ht]
          \begin{center}
   \includegraphics[height=5.5cm] {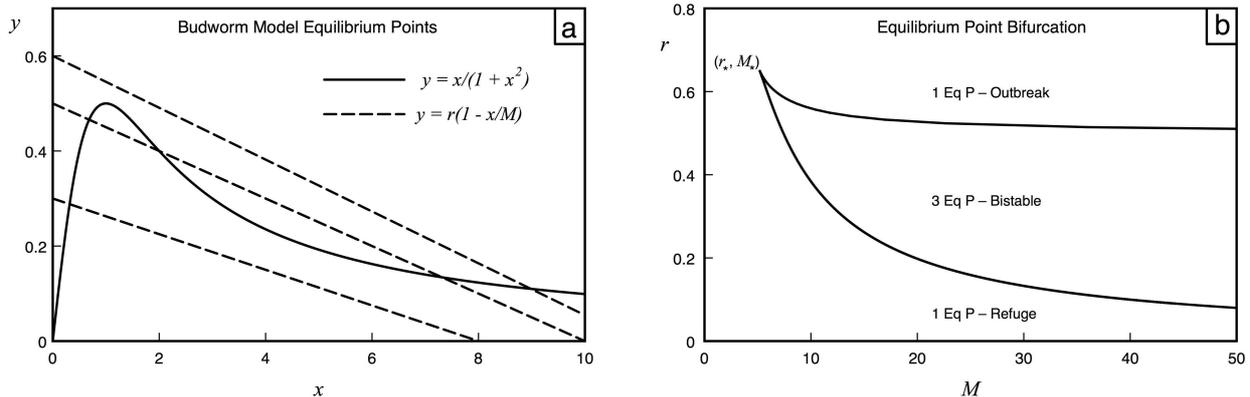}
       \end{center}
    \vspace{-20 pt}
    \caption{\small{Non-zero equilibrium points, budworm model, $\alpha=\beta=1$. (a) Graphical solution. (b) Bifurcation diagram.} } \label{fig6.fig}
 \end{figure}

Equation (\ref{budeqp.eq}) is solved graphically by comparing the intersection of the line $y=r(1-x/M)$ with the curve $y=x/(1+x^2)$, figure \ref{fig6.fig}a. If $r$ and $M$ are sufficiently small, there is only one non-zero equilibrium point, which is stable. This scenario, called  \emph{refuge}, is  where budworm numbers are kept low. Increasing the parameters leads to three equilibrium points, two of which are stable. The larger stable point is the \emph{outbreak} state where budworm numbers have been able to become large. If both parameters are high, then only the outbreak state exists. There are thus three equilibrium scenarios: refuge, outbreak and \emph{bistable} -- in which both refuge and outbreak states exist \cite{strogatz2018nonlinear}.

Bifurcation between the three equilibrium scenarios occurs when the  line $y=r(1-x/M)$ is tangent to the curve  $y=x/(1+x^2)$, i.e. when $r(1+x^2)^2=M(x^2-1)$ \cite{strogatz2018nonlinear}. Solving this equation with the equilibrium condition (\ref{budeqp.eq}) gives the bifurcation curve, figure \ref{fig6.fig}b, where:
\begin{equation}
r = \frac{2x^3}{(1+x^2)^2} \mbox{,  } \,\,\,\, M = \frac{2x^3}{(x^2-1)} \label{bifeq.eq}
\end{equation}
Setting $\dd r/ \dd x= 0$ in (\ref{bifeq.eq}) gives a critical point at $x_*=\sqrt{3}$ where $r_*=3\sqrt{3}/8$ and $M_*=3\sqrt{3}$, figure \ref{fig6.fig}b.  Thus, only one stable equilibrium state exists for either $r>r_*$ or $M<M_*$. For $M>M_*$ there are two values of $r$ at which bifurcation occurs. As $M\rightarrow \infty$, the upper bifurcation value of $r$ tends to $1/2$, with the lower one tending zero. Thus,  bistable states are more numerous for high values of population capacity $M$.

\subsection{Force Impact Analysis}
 The preceding standard equilibrium analysis of the budworm model describes the two  outcomes of the population numbers $x$, refuge or outbreak. For pest control, the desire is to keep budworm numbers under the outbreak level, either at or below the refuge level, where that exists. The equilibrium analysis has uncovered three scenarios, of which the bistable one has both stable equilibria possible,  figure \ref{fig6.fig}b. Although the existence of the equilibrium points, and their basins of attraction, informs decision making, it does not identify the points beyond which interventions fail to be effective.  The forces on the budworm numbers are now considered so that the pathways to equilibrium can be explored and thus assist in determining interventions to avoid an outbreak.
 
\begin{figure}[!ht]
          \begin{center}
   \includegraphics[height=3.8cm] {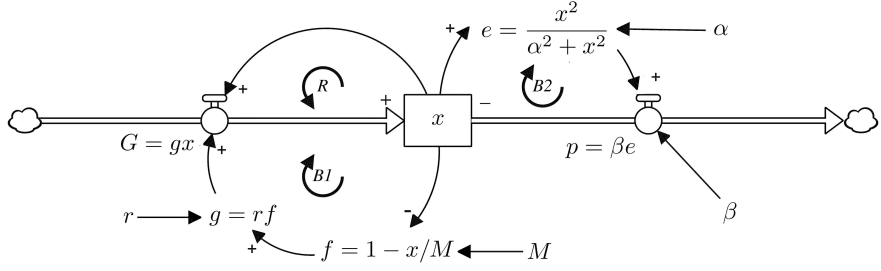}
       \end{center}
    \vspace{-20 pt}
    \caption{\small{Spruce Budworm model in stock/flow notation, with feedback loops $R$, $B_1$ and $B_2$.} } \label{fig7.fig}
 \end{figure}

The model (\ref{budworm.eq}) is expressed in stock/flow form by extending the Verhulst model, figure \ref{fig1.fig}, (\ref{verhulst.eq}), to include predation, figure \ref{fig7.fig}. The predation term $\beta x^2/(\alpha^2 + x^2)$ is treated as a single loop as it is derived from a consumption rate for the predators, whose numbers are assumed constant on the time-scale of this model \citep{ludwig1978qualitative,huisman1997formal,dawes2013derivation}. Thus there are three forces on budworm density $x$: a growth force, associated with the reinforcing feedback loop $R$; and two dissipative forces associated with capacity saturation, balancing loop $B_1$, and predation, loop $B_2$. The causally connected equation (\ref{budwormsd.eq}) indicates the pathways associated with each force. Comparing (\ref{budwormsd.eq})  with the general form of a system dynamics model (\ref{sd.eq}), $a_{11\,1} = G$, $a_{11 \,2}= fgG$ and $a_{11 \,3}= ep$. 
   \begin{equation}
 \frac{\dd x}{\dd t
} = rx_{\underline{G}}\left(1 - \frac{x_{\underline{fgG}}}{M}\right) -  \beta\frac{x_{\underline{ep}}^2}{\alpha + x_{\underline{ep}}^2} \label{budwormsd.eq}
  \end{equation}

Thus, using pathway differentiation (\ref{impact2.eq}) on (\ref{budwormsd.eq}), the impacts of the forces due to growth, saturation and predation respectively are given by:
\begin{eqnarray}
\Ip_{\mathrm{gr}}(R) &\triangleq& \Ip_{\underline{xGx}}(R)=  r  \left( 1-\frac{x}{M}\right) \label{impactbud1.eq}   \\
\Ip_{\mathrm{sat}}(B_1) &\triangleq&   \Ip_{\underline{xfgGx}}(B_1)=   - r  \frac{x}{M} \label{impactbud2.eq}  \\
 \Ip_{\mathrm{pred}}(B_2)  &\triangleq& \Ip_{\underline{xepx}}(B_2) = -\frac{2\alpha^2\beta x}{(\alpha^2 + x^2)^2} \label{impactbud3.eq}
 \end{eqnarray}
 As in the Verhulst model, the absolute values of the impacts associated with the growth and capacity saturation forces, (\ref{impactbud1.eq}--\ref{impactbud2.eq})  are  monotonically decreasing and increasing respectively. However, the predation force (\ref{impactbud1.eq}) has maximum absolute impact when the population has a specific value $x_{p}$ (\ref{xsat.eq}). Therefore, for higher population values, predation has a diminishing effect compared with capacity saturation and is thus less able to exert the control needed to avoid a budworm outbreak.  The population value for which the predation and saturation impacts (\ref{impactbud2.eq}--\ref{impactbud3.eq}) are equal is given by $x_s$:
 \begin{equation}
x_p = \frac{1}{\sqrt{3}}, \,\,\,\,\, x_{s}=\sqrt{\sqrt{\frac{2M}{r}}-1} \,\,\,\,\,\, \mbox{for}\,\,\,\,\, \alpha=\beta=1\label{xsat.eq}
 \end{equation}
   Once budworm numbers exceed $x_s$, the predation force falls below that of saturation and does not recover.

\begin{figure}[!ht]
          \begin{center}
   \includegraphics[height=5.5cm] {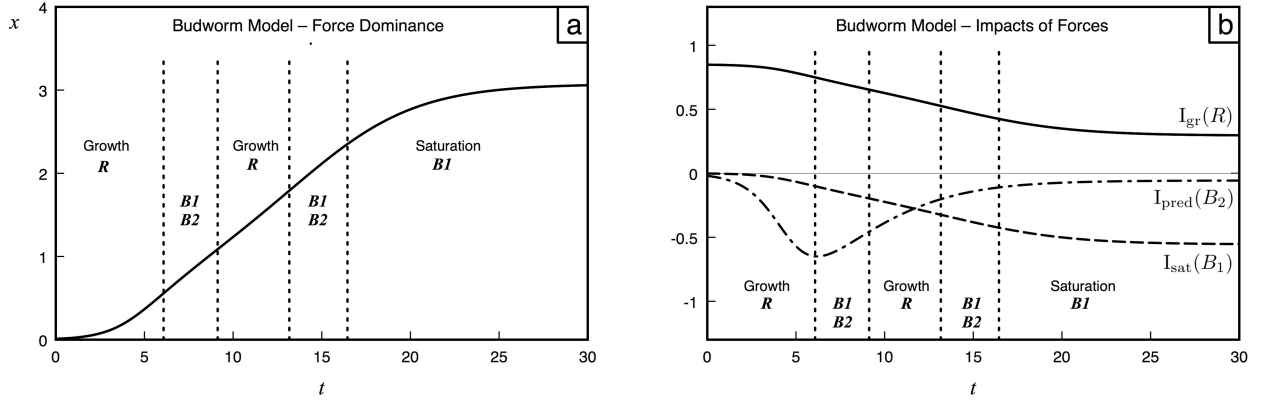}    
       \end{center}
    \vspace{-20 pt}
    \caption{\small{(a) Solution to Spruce Budworm model for population $x$ and regions of force dominance associated with loops $R$, $B_1$, $B_2$; $r=0.85$, $M=4.7$, $\alpha=\beta=1$ and $x_0=0.01$. (b) Impacts of forces: $\Ip_{\mathrm{gr}}(R),\Ip_{\mathrm{sat}}(B_1)$, $\Ip_{\mathrm{pred}}(B_2)$. } }  \label{fig8.fig}
 \end{figure}

A typical growth scenario is given in figure \ref{fig8.fig}, where the parameter values have set the equilibrium state at outbreak (figure \ref{fig6.fig}b). The periods of force dominance are indicated on both population and impact graphs, (figures \ref{fig8.fig}a and \ref{fig8.fig}b). Where behaviour is decelerating, but neither balancing loop  is  sufficient alone to exceed the impact of the growth loop $R$, then, following \citet{hayward2014model}, the minimum dominant set of impacts is used,  indicated by $B_1B_2$\footnote{Where more than two impacts of the same polarity are compared, the minimum set of impacts with the largest values is used to explain dominance \citep{hayward2014model}, described by \citet{sato2016state} as a sufficient but not necessary set.}.

 After the initial acceleration, dominated by $\Ip_{\mathrm{gr}}(R) $, there follows a long period of steady growth where dominance switches from $\Ip_{\mathrm{sat}}(B_1)+ \Ip_{\mathrm{pred}}(B_2)$ to the growth process and back again, figure \ref{fig8.fig}a. Equilibrium is finally achieved by the saturation force. Although the  predation force has a larger impact than that of saturation in the first period of combined balancing growth, $B_1B_2$ figure \ref{fig8.fig}b, it has already exceeded its maximum absolute value as $x>x_p$, and it is having less effect in slowing budworm growth. During the second growth period, the predation force falls below that of saturation as $x>x_s$ so that in the second period  of combined balancing impacts that follows, $B_1B_2$, predation is having a negligible effect and budworm numbers head to the outbreak equilibrium. The results suggest that an intervention to restrict budworm numbers and avoid an outbreak should be applied before the predation force reaches its maximum impact. If the intervention is delayed, the assistance of the predators is diminished, and pest control would become much harder.
 
The scenario in figure \ref{fig8.fig} is one of many possibilities that occur with changes in the parameters $r$ and $M$. A consideration of the forces, along with the equilibrium analysis figure \ref{fig6.fig}, identifies the categories of behavioural scenarios. The concept of bifurcation is extended to the transitions between reinforcing and balancing forces,  which occur at the inflexion points of the curve, $\ddot{x}=0$. These will be referred to as \emph{impact transition points}, as they mark the change in polarity of the net force impact on the stock $x$, the sum of the impacts (\ref{impactbud1.eq}--\ref{impactbud3.eq}).

Setting the time derivative of  (\ref{budworm.eq}) to zero gives the equation of the impact transition points:
      \begin{equation}
      r  \left( 1-\frac{2x}{M}\right) =\frac{2 x}{(1 + x^2)^2} \label{itpeq.eq}
      \end{equation}
      
\begin{figure}[!ht]
          \begin{center}
   \includegraphics[height=5.5cm] {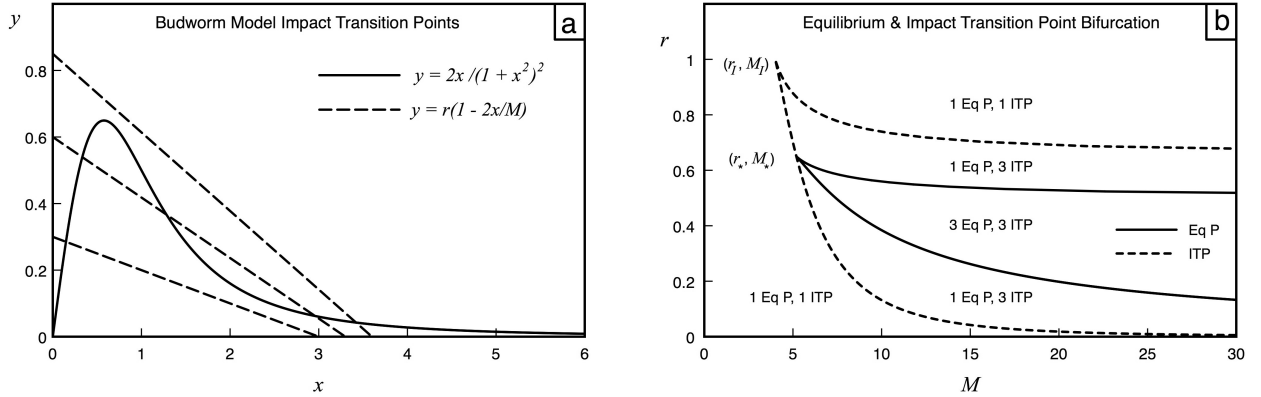}
       \end{center}
    \vspace{-20 pt}
    \caption{\small{Impact transition points (ITP), Spruce Budworm model, $\alpha=\beta=1$. (a) Graphical solution. (b) Bifurcation diagram} } \label{fig9.fig}
 \end{figure}

      As in the case of the equilibrium points, (\ref{itpeq.eq}) is solved graphically by comparing the line $y=r(1-2x/M)$ with the curve $y=2x/(1+x^2)^2$, figure \ref{fig9.fig}a. There are scenarios where there are either one, two or three impact transition points. Bifurcation occurs when the line is tangent to the curve, the case of two impact transition points: $ r/M  = (3x^2-1)/(1+x^2)^3$. Solving this equation with (\ref{itpeq.eq}) gives the bifurcation curve  for the impact transition points, figure \ref{fig9.fig}b, where:
      
       \begin{equation}
r=\frac{8x^3}{(1+x^2)^3}  \mbox{   and   }   M = \frac{8x^3}{3x^2-1} \label{itpbif.eq}
 \end{equation}
      
 Setting $\dd r / \dd x=0 $ in (\ref{itpbif.eq}) gives a critical point for impact transitions at $x_I=1$, $r_I=1$ and $M_I=4$. Thus, only one impact transition, i.e. change of loop dominance, exists for $r>r_I$ or $M<M_I$. As $M\rightarrow \infty$, the upper bifurcation value of $r$ tends to $r_*$, the critical value of the equilibrium bifurcation. Comparing both types of bifurcations, figure \ref{fig9.fig}b, shows that the bistable states, i.e. those with three equilibrium points, also have three impact transition points as expected. However, the predation and saturation states, i.e. those with 1 equilibrium point, are subdivided according to the number of impact transition points. Thus, the following growth scenarios are identified:
 
 \begin{enumerate}
\item Low Capacity. For $M<M_I=4$ there are no bifurcations in $r$. 

\item Moderate Capacity. For $M_I=4<M<M_*=3\sqrt{3}$ there are no equilibrium point bifurcations in $r$ but there are two impact transition point bifurcations. Thus, three cases: Low growth $r<r_{i-}$;  moderate growth $r_{i-}<r<r_{i+}$ and high growth $r_{i+}<r$, where $i-$ and $i+$, refer to the lower and upper impact transition bifurcation curves respectively.

\item High Capacity. For $M_*=3\sqrt{3}<M$ there are bifurcations  in $r$ for both types of points. Thus, five cases: Very low growth $r<r_{i-}$; low growth $r_{i-}<r<r_{e-}$; moderate growth $r_{e-}<r<r_{e+}$; high growth $r_{e+}<r<r_{i+}$ and very high growth $r_{i+}<r$, where $e-$ and $e+$, refer to the lower and upper equilibrium  bifurcation curves respectively.
\end{enumerate} 

\begin{figure}[!ht]
          \begin{center}
   \includegraphics[height=5.5cm] {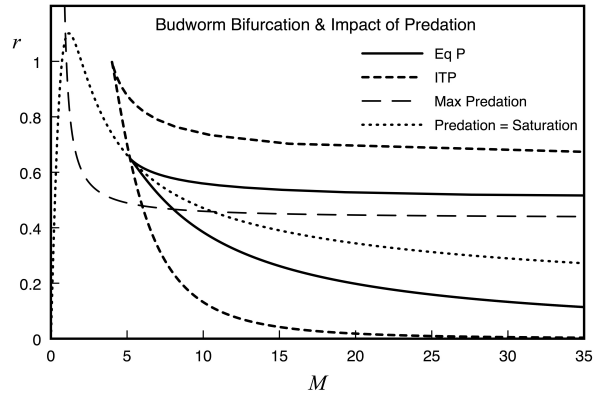}
       \end{center}
    \vspace{-15 pt}
    \caption{\small{Equilibrium point (Eq~P) and impact transition point (ITP) bifurcation diagram for Spruce Budworm model, with curves where an equilibrium point equals $x_p$ (maximum predation impact) and $x_s$  (equal saturation and predation impacts).} } \label{fig10.fig}
 \end{figure}

Each of these nine scenarios will have a different balance of forces as budworm numbers approach equilibrium. The relative impacts of predation and saturation can be clarified by comparing, in each scenario, the bifurcation regions with the maximum predation impact points $x_p$, and the equal saturation and predation impact point $x_s$ (\ref{xsat.eq}).  The parameter values where these points equal an equilibrium point are displayed on the bifurcation diagram, figure \ref{fig10.fig}. The results show that it is the lower growth scenarios where predation impact is of the most assistance to control as it remains higher than saturation and does not pass its maximum. The high capacity, moderate growth scenario has equilibria where predation is yet to achieve its maximum, even though the saturation equilibrium is possible. This suggests that early intervention to strengthen the action of the predators could avoid the saturation outbreak.  
 
\begin{figure}[!ht]
          \begin{center}
   \includegraphics[height=11.3cm] {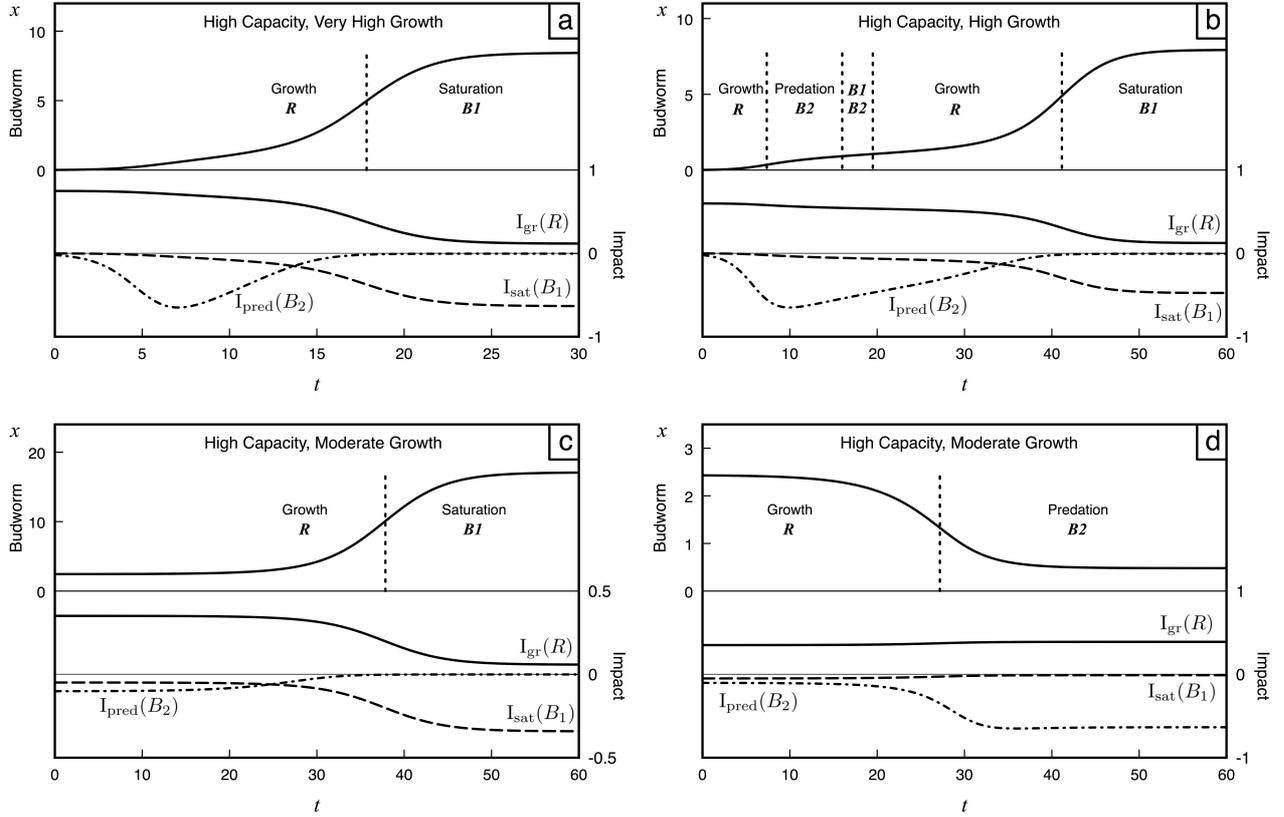}
       \end{center}
    \vspace{-20 pt}
    \caption{\small{Solution to Spruce Budworm model for population $x$ and regions of force dominance for high capacity scenario. (a) Very high growth, $r=0.75$, $M=10$, $x_0=0.01$. (b) High growth, $r=0.6$, $M=10$, $x_0=0.01$. (c) Moderate growth $r=0.4$, $M=20$, $x_0=2.44$. (d) $r=0.4$, $M=20$, $x_0=2.43$. }} \label{fig11.fig}
 \end{figure}

 Six scenarios are now considered that illustrate the benefits of the force concept.
In the high capacity, very high growth case, figure \ref{fig11.fig}a, the predation impact is much higher than saturation, which helps slow the initial growth in budworm numbers. Even in this extreme scenario, rapid acceleration is delayed. In the high growth scenario, figure \ref{fig11.fig}b, there is a long initial period where the growth process is slowed due to the high impact of the predators. Thus, although both scenarios result in the saturation equilibrium, an early intervention that strengthens predator action could hold budworm numbers low.

In the high capacity moderate growth scenario, figures \ref{fig11.fig}c--d, both stable equilibria are possible depending on the position of the initial value $x_0$ relative to the unstable equilibrium. The figures show the case in which predator impact 
initially exceeds that of saturation. If budworm numbers start above the unstable equilibrium, saturation eventually exceeds predation, but only after an extended period, figure \ref{fig11.fig}c. If an intervention tipped numbers to fall below the unstable point, predation impact remains the largest, figure \ref{fig11.fig}d. Again, results suggest interventions should assist predator action.

\begin{figure}[!ht]
          \begin{center}
   \includegraphics[height=5.5cm] {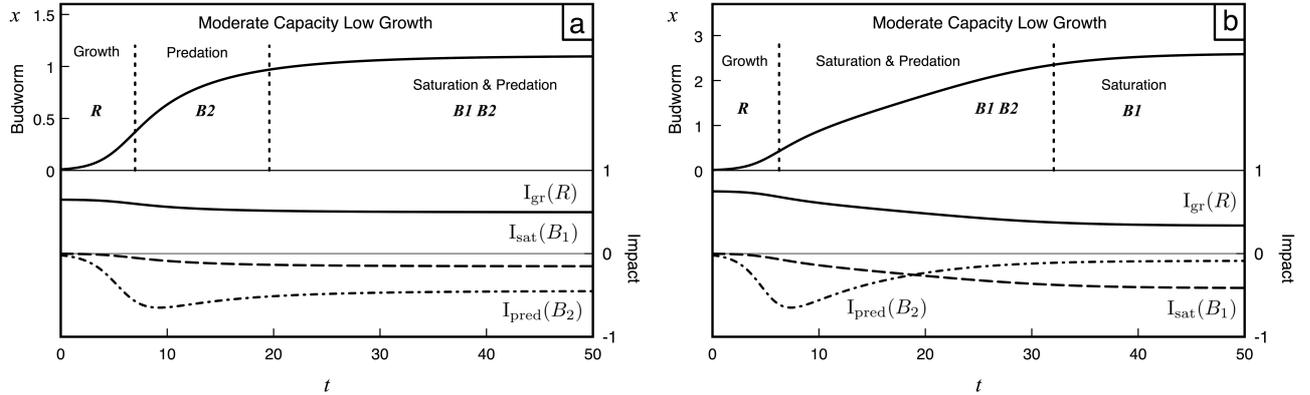}
       \end{center}
    \vspace{-20 pt}
      \caption{\small{Solution to Spruce Budworm model for population $x$ and regions of force dominance for moderate capacity, low growth scenario, $M=4.7$, $x_0=0.01$. (a)  $r=0.65$, $x_{eq}>x_s$. (b) $r=0.75$, $x_{eq}<x_s$.  }}  \label{fig12.fig}
 \end{figure}

In the moderate capacity scenarios, there is only one stable equilibrium point $x_{eq}$, figure \ref{fig9.fig}b. However, in the low growth case, it is a combination of predation and saturation impacts that achieves equilibrium. For values of $r$ where equilibrium is achieved before predation impact falls below that of saturation, figure \ref{fig10.fig}, it is predation force that initially slows the growth, only later being joined by saturation, figure \ref{fig12.fig}a. For large $r$, where predation impact drops below saturation before equilibrium, there is a long period where the predation force assists saturation in bringing about equilibrium, figure \ref{fig12.fig}b. Even in the case of moderate growth, the predation force still plays a significant role in achieving equilibrium, figure \ref{fig8.fig}. Similar results can be shown for the low capacity scenario.

Thus, a comparison of the forces, using impact as a measure, has shown the importance of the role of predators in slowing growth in a range of growth and capacity scenarios, and can thus  inform control policies for potential budworm outbreaks.

 \pagebreak

  \section{Predator-Prey Model} \label{predprey}
   \subsection{Model and Equilibrium Analysis}
   A Lotka-Volterra system with a carrying capacity on the predator \cite{murray2002mathematical,turchin2003complex,braun1983differential} will now be considered, both as an example of a higher order model, and to demonstrate that the force concept provides a physical explanation of stability. Let a prey, numbers  $x$, have constant  birth rate $a$ and death rate $by$, which is proportional to predator numbers $y$ (\ref{lv1.eq}). The predator grows at a rate according to prey number $cx$ and is subject to a death rate $d+ey$ that increases with predator numbers due to environmental constraints (\ref{lv2.eq}).
 \begin{eqnarray}
\frac{\dd x}{\dd t} &=& ax-bxy \label{lv1.eq} \\
\frac{\dd y}{\dd t} &=&cxy - (d + ey)y \label{lv2.eq}
\end{eqnarray}
The parameters can be reduced to two by the transformations $x \rightarrow dx/c$, $y \rightarrow dy/b$ and $t \rightarrow t/d$, with appropriate redefinitions for $a$ and $e$. Thus it assumed $b=c=d=1$. 

Setting (\ref{lv1.eq}--\ref{lv2.eq}) to zero determines  the equilibrium points with their stability computed using the system Jacobian:
\[
J = \left[   
 \begin{array}{cc}
  a-y  & -x \\
   y  &   x -1-2ey
  \end{array}
\right]
 \]
There are two physical equilibrium points: $(0,0)$, which is always unstable; and $(1+ ae, a)$ which is stable for $e>0$. For $e >e_* \triangleq 2 + 2\sqrt{1+1/a} $ the non zero equilibrium point is a stable node. For $0<e<e_*$, the point is a stable focus, i.e. damped oscillations. For $e=0$, this non-zero point has neutral stability exhibiting closed path oscillations dependent on initial conditions -- the classic Lotka-Volterra model \cite{murray2002mathematical}.

\subsection{Force Impact Analysis}
\begin{figure}[!ht]
          \begin{center}
   \includegraphics[height=6.5cm] {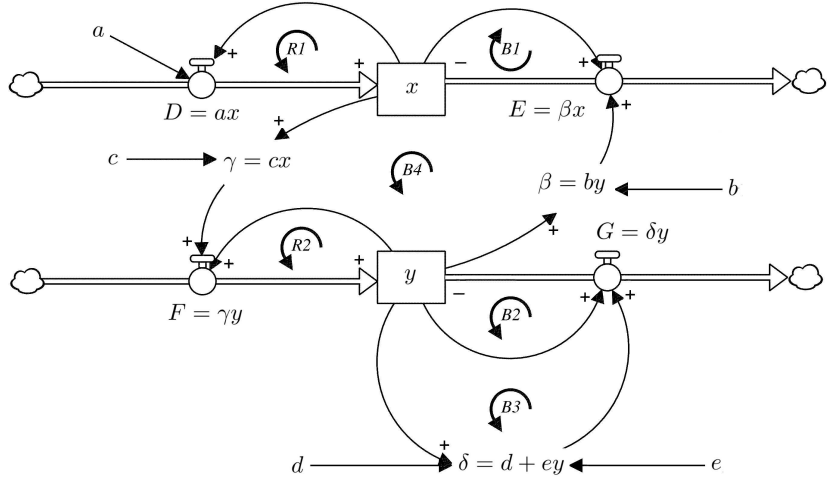}
       \end{center}
    \vspace{-15 pt}
    \caption{\small{Predator-Prey model in stock/flow notation, with feedback loops $R_1$, $B_1$, $R_2$, $B_2$, $B_3$ and $B_4$.  }} \label{fig13.fig}
 \end{figure}

The predator-prey model is expressed in stock/flow form in figure \ref{fig13.fig}, with corresponding causally connected differential equations (\ref{lv5.eq}--\ref{lv6.eq}). There are three forces on prey numbers $x$: a growth force due to births, associated with the reinforcing loop $R_1$; a dissipative force due to deaths, associated with loop $B_1$; and a force from $y$, associated with consumption by predators, which is part of the second-order predation loop $B_4$. There are four forces on predator numbers $y$: growth due to births, associated with loop $R_2$; a force from $x$, associated with the benefits of consuming prey, which is the other part of the predation loop $B_4$; and two dissipative forces, associated with loops $B_2$, deaths in the absence of capacity effects, and $B_3$, additional deaths due to environmental saturation. Comparing (\ref{lv5.eq}--\ref{lv6.eq})  with the general form of a system dynamics model (\ref{sd.eq}), $a_{11\,1} = D$, $a_{11 \,2}= E$, $a_{21\,1} = \beta E$, $a_{12\,1} = \gamma F$, $a_{22\,1} = F$, $a_{22 \,2} = G$ and $a_{22 \,3}= \delta G$. 
  \begin{eqnarray}
\frac{\dd x}{\dd t} &=& ax_{\underline{D}}-bx_{\underline{E}}y_{\underline{\beta E}} \label{lv5.eq} \\
\frac{\dd y}{\dd t} &=&cx_{\underline{\gamma F}}y_{\underline{F}} - \left(d + ey_{\underline{\delta G}}\right)y_{\underline{G}} \label{lv6.eq}
\end{eqnarray}

Thus, using pathway differentiation (\ref{impact2.eq}) on (\ref{lv5.eq}--\ref{lv6.eq}), the impacts of the forces  are given by:
\begin{eqnarray}
\Ip_{\mathrm{gr}\,x}(R_1) &\triangleq& \Ip_{\underline{xDx}}(R_1)=  a\label{impactpp1.eq}   \\
\Ip_{\mathrm{dis}\,x}(B_1) &\triangleq&   \Ip_{\underline{xEx}}(B_1)=   - by \label{impactpp2.eq}    \\
\Ip_{\mathrm{gr}\,y}(R_2) &\triangleq& \Ip_{\underline{yFy}}(R_2)=cx \label{impactpp3.eq}\\
\Ip_{\mathrm{dis}\,y}(B_2) &\triangleq&\Ip_{\underline{yGy}}(B_2)=-(d+ey) \label{impactpp4.eq}\\
\Ip_{\mathrm{sat}\,y}(B_3) &\triangleq& \Ip_{\underline{y\delta Gy}}(B_3)=-ey \label{impactpp5.eq} \\
\Ip_{\mathrm{pred}\,x}(B_4) &\triangleq&    \Ip_{\underline{y\beta Ex}}(B_4)= -\frac{by(cx - d-ey) }{a  - by} \label{impactpp6.eq} \\
\Ip_{\mathrm{pred}\,y}(B_4) &\triangleq& \Ip_{\underline{x\gamma Fy}}(B_4)=  \frac{cx(a  - by)}{(cx - d-ey)} \label{impactpp7.eq} 
 \end{eqnarray}
 The impacts associated with the first-order loops (\ref{impactpp1.eq}--\ref{impactpp5.eq}) are the loop gains. Of these, all except $\Ip_{\mathrm{gr}\,x}(R_1)$  depend on the stock values due to the non-linearities of the model. The predation impacts (\ref{impactpp6.eq}--\ref{impactpp7.eq}) potentially have singularities, which will allow these forces to change polarity. The predation loop gain, $\Ip_{\mathrm{pred}\,x}(B_4)\Ip_{\mathrm{pred}\,y}(B_4)=\Ip_{\underline{y\beta Ex}}(B_4)\Ip_{\underline{x\gamma Fy}}(B_4)=-bcxy$ is variable but always has negative polarity.

\begin{figure}[!ht]
          \begin{center}
   \includegraphics[height=5.5cm] {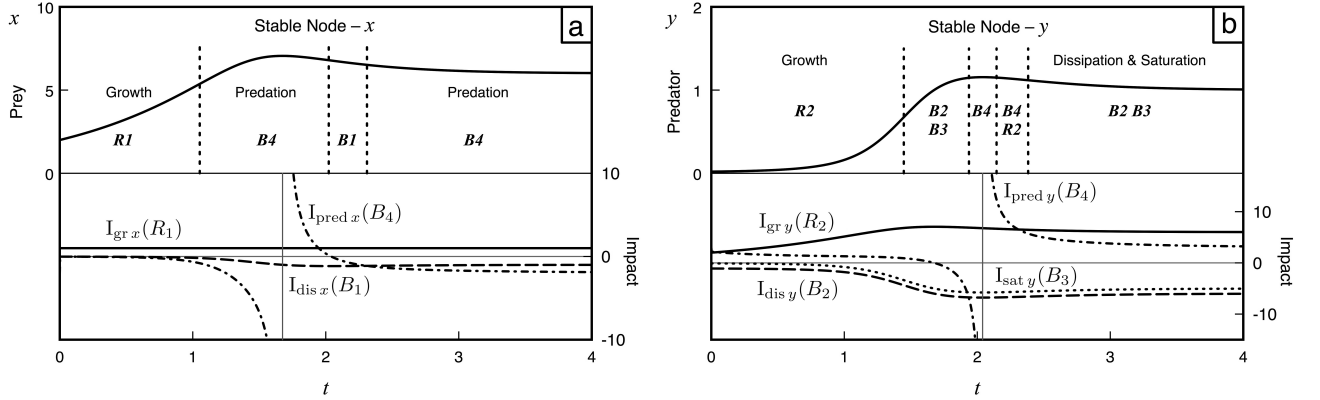}    
       \end{center}
    \vspace{-20 pt}
    \caption{\small{Solution to predator-prey model for populations $x$, $y$ and regions of force dominance with stable node, $a=b=c=d=1$, $e=5>e_*$ and $(x_0,y_0) = (2,0.02)$. $B_4$ is associated with the predation forces $\Ip_{\mathrm{pred}\,x}(B_4)$, $\Ip_{\mathrm{pred}\,y}(B_4)$.  } }  \label{fig14.fig}
 \end{figure}

For the case of the stable node,  $e>e_*$,  predation $B_4$ is the dominant force controlling the growth of the prey $x$ and bringing it to equilibrium, with the first-order dissipation process, $B_1$, only playing a minor role, figure \ref{fig14.fig}a. The predation force shows the characteristic change of polarity in its impact $\Ip_{\mathrm{pred}\,x}(B_4)$ on $x$ at the turning point,  along with a momentarily infinite value. Thus the action of the predator $y$ in consuming prey is solely responsible for changing prey growth to decline. Likewise, predation is also responsible for turning growth into decline for the predator, figure \ref{fig14.fig}b. However, predation plays no role in achieving predator equilibrium as it has a positive impact in the final phase to match the negative impact of predation on the prey (figure \ref{fig14.fig}a). Instead, it is a combination of dissipation and saturation that brings the predator to equilibrium, similar to results in models of  competition and cooperation \cite{simin2018dynamical}. 

Force dominance at equilibrium is a general result. Applying the stable equilibrium point to the  impacts (\ref{impactpp1.eq}--\ref{impactpp2.eq}) gives  $\Ip_{\mathrm{gr}\,x}(R_1) = a = -\Ip_{\mathrm{dis}\,x}(B_1)$; thus they cancel identically. Using linearisation, (see \ref{appendix2}), the predation impact on $x$  (\ref{impactpp6.eq}) is $\Ip_{\mathrm{pred}\,x}(B_4) = -ae/2 + \sqrt{a^2e^2-4a-4a^2e}/2$  at equilibrium, which is less than $-a$ for all $e>0$. Thus, $|\Ip_{\mathrm{pred}\,x}(B_4)|>| \Ip_{\mathrm{dis}\,x}(B_1)|$, showing that the predation force is responsible for prey equilibrium, for $e>e_*$. For the predator  at equilibrium, from (\ref{impactpp3.eq}--\ref{impactpp4.eq}), $\Ip_{\mathrm{gr}\,y}(R_2) = 1+ae=-\Ip_{\mathrm{dis}\,y}(B_2)$. Thus, $|\Ip_{\mathrm{sat}\,y}(B_3)|>|\Ip_{\mathrm{pred}\,y}(B_4) |$ to give a net negative impact for stability. Therefore, it takes both negative impact forces, dissipation and saturation, to bring $y$ to equilibrium.

\begin{figure}[!ht]
          \begin{center}
   \includegraphics[height=5.5cm] {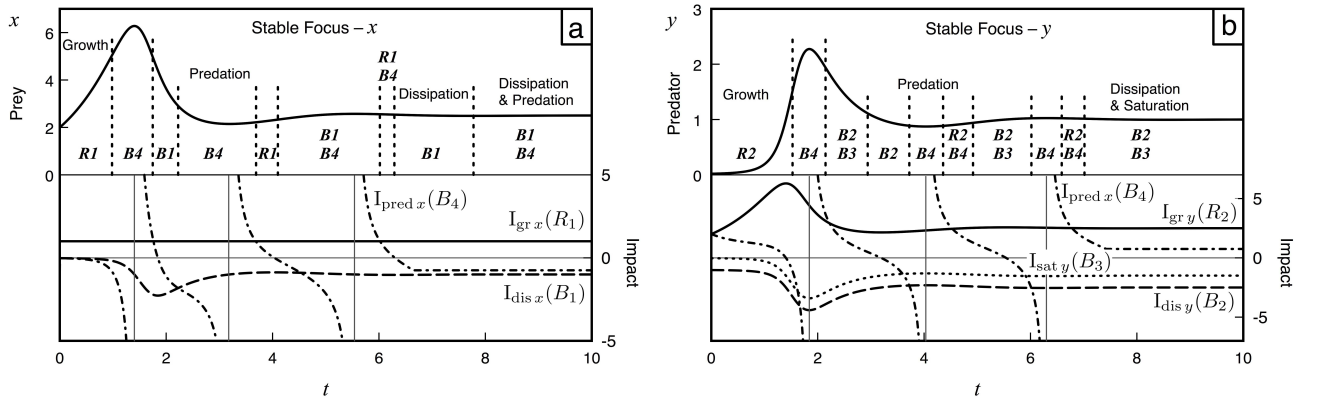}    
       \end{center}
    \vspace{-20 pt}
    \caption{\small{Solution to predator-prey model for populations $x$, $y$ and regions of force dominance with stable focus, $a=b=c=d=1$, $e=1.5<e_*$ and $(x_0,y_0) = (2,0.02)$.   } }  \label{fig15.fig}
 \end{figure}

For the case of the stable focus, $0<e<e_*$, the cyclical behaviour of $x$ is explained by repeated periods of $R_1, B_4, B_1, B_4$, with predation controlling the change from growth to decline and vice versa, figure \ref{fig15.fig}a. However, as oscillations become damped the growth process, $R_1$, requires assistance from predation, now with positive impact due to falling predator numbers, about $t=6$ figure \ref{fig15.fig}a. For determining dominance, as damping becomes insignificant, the impact of the second-order predation loop is replaced with its average value over one cycle $\langle\Ip_{\mathrm{pred}\,x}(B_4)\rangle=-ae/2$ (\ref{appendix2}). This eliminates the repeated infinite impact values, which are an artefact of the ratio measure of impact. For $y$, the cycles are $R_2, B_4, B_2  B_3, B_2, B_4$, with increasing instances of multiple loop dominance as oscillations are damped, figure \ref{fig15.fig}b. As in the case of the stable node, equilibrium in $y$ is brought about by predator dissipation and saturation together (replacing predation impact by its average value over one cycle $\langle \Ip_{\mathrm{pred}\,y}(B_4)\rangle=ae/2$, \ref{appendix2}). By contrast with the node case, the prey will need dissipation to assist  predation to achieve equilibrium when $|\Ip_{\mathrm{dis}\,x}(B_1)|>|\Ip_{\mathrm{pred}\,x}(B_4)|$ i.e. if $e<2$. 

\begin{figure}[!ht]
          \begin{center}
   \includegraphics[height=5.5cm] {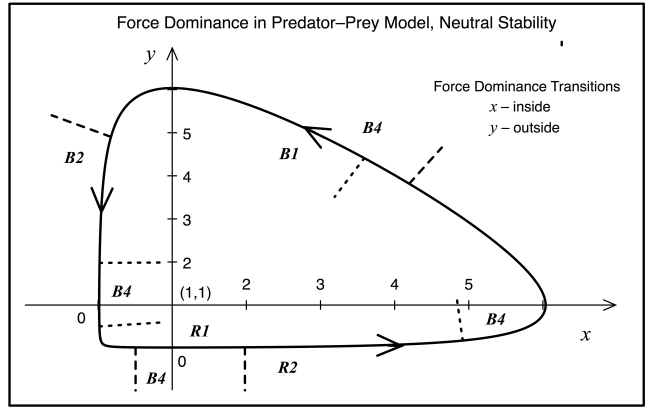}
       \end{center}
    \vspace{-15 pt}
    \caption{\small{Solution to predator-prey model for populations $x$, $y$ and regions of force dominance with neutral stability, $a=b=c=d=1$, $e=0$ and $(x_0,y_0) = (2,0.02)$.} Axes centred on equilibrium point (1,1). Changes of force dominance for prey $x$ are indicated inside the phase path; those for predator $y$ are indicated outside the phase path.} \label{fig16.fig}
 \end{figure}

The case of neutral stability, $e=0$, is the classic Lotka-Volterra model and exhibits path-dependent closed cycles \cite{murray2002mathematical}. Figure \ref{fig16.fig} shows the regions of force impact dominance superimposed on the phase plane where the change of the impact dominance of $x$ is on the inside of the closed curve and that of $y$ is on the outside.  The cycle starts with both populations growing and accelerating, $R_1$ and $R_2$ dominance, until the predation $B_4$ impact slows $x$ and causes its decline.  The predator $y$ moves from acceleration to predation dominance as it reacts to the dwindling food supply, quickly followed by dissipation dominance in $x$ as its numbers continue to collapse.  Eventually the predation force causes the predator to decline with the dissipation forces, $B_1$ and $B_2$, leading to the collapse of both populations. Once the predator numbers are sufficiently small, the prey is able to recover, first through the positive effects of predation $B_4$, then through $R_1$.

Over one cycle,  the average values of the populations are the equilibrium values \cite{braun1983differential}. Thus, as the first-order impacts are either linear in population numbers, or constant, the average impacts over one cycle balance for each population are: $\langle\Ip_{\mathrm{gr}\,x}(R_1)\rangle =-\langle\Ip_{\mathrm{dis}\,x}(B_1)\rangle =a$, $\langle\Ip_{\mathrm{gr}\,y}(R_2)\rangle =-\langle\Ip_{\mathrm{dis}\,y}(B_2)\rangle =1$. From (\ref{impactpp6.eq}) the average value between $t_1$ and $t_2$ of the predation impact on $x$   is
\[
\langle\Ip_{\mathrm{pred}\,x}(B_4)\rangle = \frac{1}{t_2-t_1}\int_{t_1}^{t_2}\Ip_{\mathrm{pred}\,x}(B_4) \dd t =  \left[\frac{\ln|a-y|}{t_2-t_1}\right]_{y(t_1)}^{y(t_2)}
\]
Over one cycle, $y(t_1)=y(t_2)$, thus $\langle\Ip_{\mathrm{pred}\,x}(B_4)\rangle$ =0. Likewise, $\langle\Ip_{\mathrm{pred}\,y}(B_4)\rangle = 0$.    Therefore, the undamped cyclical behaviour of the Lotka-Volterra model can be explained by the net balance of forces on each population stock.  




 In summary,  the use of the concept of force has explained the oscillations of the Lotka-Volterra model  by a lack of sufficient dissipation. Although dissipation is present in the form of deaths of prey and predators, these are, on average, cancelled by the birth processes. Only when there is additional dissipation due to environmental carrying capacity effects on the predator does the net balance of forces produce stability.

 \section {Discussion and Conclusion}
 This paper uses the concept of force, introducing it into population dynamics.  Its explanatory power for the understanding of model behaviour is demonstrated. The influence exerted by one model variable on another is understood as a force, operating within feedback loops implied by the model, providing a narrative explanation of the curvature in variable behaviour. The behaviour of variables in ordinary differential equation models is interpreted in terms of the balances of forces they exert on each other, through use of the concept of loop impact of \citet{hayward2014model}, and the Newtonian Interpretative Framework of \citet{hayward2017newton}. In order to locate the forces within feedback loops, and thereby employ the framework, the models are re-expressed in system dynamics form using the stock/flow notation of \citet{forrester1961industrial}. The loop dominance method of \citet{hayward2014model} is used to determine the force, or forces, responsible for the curvature in variable behaviour at any point in time. The methodology described in this paper offers an enhancement to standard equilibrium analysis. An additional viewpoint is provided by force impact, the ratio of a force's acceleration of a state variable to the rate of change of the variable. This approach is useful in the analysis of models for which behaviour is better understood by identifying the transient phases, achieved here in mechanical terms. It is noted that impact, as defined in this paper, is independent of variable units, and it measures the effect on the state variables contained in its associated feedback loop. The paper also develops a symbolic notation to encapsulate the network of forces in population models,  thus enabling differentiation along a given causal pathway to determine the impact of the forces.

 The Newtonian Interpretive Framework is applied to two well-known models: the Spruce Budworm model of \citet{ludwig1978qualitative} and an extended predator-prey model. In each case, an analysis of the effect of the forces is used to enhance the standard equilibrium analysis as an explanation of behaviour. In the case of the Spruce Budworm model, the force analysis produces a bifurcation diagram for impact transitions, analogous to that determined by equilibrium analysis, figures \ref{fig9.fig}b and \ref{fig10.fig}. This provides a richer explanation of behavioural scenarios as the analysis now addresses the whole of the transient phase, not just final equilibrium. Whereas the equilibrium analysis suggests the need for interventions, the force analysis adds far greater clarity as to when those interventions should occur. For the predator-prey model, the force analysis shows that Lotka-Volterra cycles occur because the dissipative effects of deaths and predation are insufficient to exceed the growth forces. The additional dissipation of environmental effects is required to produce stability. A further contribution made to the analysis, by a consideration of forces, is the replacement of oscillations by their net force impact. For each model, the consideration of forces provides a more natural narrative embedded in the real world of the processes affecting the populations, rather than one reliant on the mathematical concept of stability and its classification.

A key feature of the Newtonian Interpretive Framework is that force is described in the Newtonian sense of causing acceleration in variables, section \ref{force.sec}. Thus, Newton's laws of motion can be identified in models based on systems of ordinary differential equations once they are expressed in second-order form, equations (\ref{jacobi.eq}) and (\ref{sddiff.eq}) \citep{hayward2017newton}. These equations express Newton's second law for each variable in terms of the effects of dependent variables. The absence of dependencies expresses the first law of motion, $\dd^2 x_i /\dd t^2=0$ \footnote{An example of Newton's third law is given in \citet{hayward2017newton}}. This contrasts with other definitions of force in dynamical systems. For example, \citet{montroll1978social} defines force as a deviation from exponential behaviour. Thus his ``first law of social dynamics'' describes the absence of any social, economic or ecological force as the standard growth model   $\dd (\ln x) / \dd t = c$, ($c$ constant),    which can be expressed as $\dd^2 (\ln x)/\dd t^2 = 0$.  Therefore, he views population in the Verhulst model as being subject to a single force deviating the population from exponential behaviour, as does \citet{ausloos2013another} who modified and applied Montroll's work. The Newtonian Interpretive Framework views the population in the Verhulst model in terms of two competing forces (\ref{verhurlstimpact.eq}), figure \ref{fig2.fig}.  Extensions to the model, as in the Spruce Budworm and predator-prey models (sections \ref{spruce.sec} and \ref{predprey}), introduce additional forces. Thus, the exponential growth process is seen as a force due to natural population growth which competes with ecological and predation forces. In the framework, the exponential process is characterised by constant impact, a feature alluded to by \citet{montroll1978social}, though not in force terms. The treatment of the growth process as a force in the Newtonian Interpretive Framework provides a direct analogy with Newton's laws of motion and allows the concept of force presented here to be understood in the orthodox Newtonian sense.

The Newtonian Interpretive Framework can also be compared with the approach of biological feedback modelling \citep{thomas1990biological,cinquin2002roles,deangelis2012positive} as both explain behaviour using the competing effects of positive and negative feedback.  However, the Newtonian Interpretive Framework, being based on Forester's system dynamics, is able to identify forces with each individual pathway of influence, even when there are multiple pathways between the same variables. Biological feedback models, whose emphasis is primarily one of equilibrium and control, generally aggregate multiple pathways, and are thus unable to  develop the concept of force. It is noted that some authors, e.g. \citet{cinquin2002roles}, use interaction graphs that capture individual causal pathways, suggesting that biological feedback models have the potential for a Newtonian interpretation. The Newtonian Interpretive Framework also suggests conjectures, similar to those proposed in 
biological feedback theory.  From this paper, and previous work \citep{hayward2014model,hayward2017newton},  it is conjectured that no model is stable unless there is at least one dissipative force associated with a first-order balancing loop. This is similar to the conjectures  of Thomas \citep{thomas1990biological, gouze1998positive} which have been addressed by formal proof using graph theory. Categorisation of models in terms of the presence of types of forces would assist the narrative of the Newtonian Interpretive Framework by providing simple explanations of model behaviour.

The work presented in this paper demonstrates  the applicability of Newtonian concepts to population dynamics, in particular that of force. There is much scope for further research. Using an analysis of force impact in a wider range of models would determine the extent to which the mechanical analogy aids the understanding of  model behaviour. While this paper has addressed population models in the domain of biology, it is the authors' intention that the approach is applied to other domains, such  as sociology, economics and politics. The paper highlights the key benefits of the consideration of force. The standard equilibrium analysis of models is enhanced through consideration of force
 impact, enabling an explanation of behaviour across the whole time period, transient phases as well as equilibrium. The location, by the method, of transitions of force dominance enables the identification of behavioural scenarios, which in turn improve
 the depth and scope of model analysis. Additionally, impact as a measure of force enables a Newtonian approach to analysis to be related to the feedback approach. Lastly, the authors suggest that the Newtonian Interpretive Framework, through its recognition
 of the mechanical nature of ordinary differential equation models, explains model behaviour in terms familiar to broader audiences, enhancing the communication of models and their results.  The framework proposed therefore makes a contribution to sociophysics, providing a theoretical framework for a form of ``sociomechanics''.

    \section*{References}

\bibliographystyle{unsrtnat}   
\bibliography{PHYSA19247Hayward}

 \clearpage
  \appendix
  
 \section {Causal Pathway Notation} \label{appendix1}
In a system dynamics model, cause and effect are represented by a set of equations such as 
(\ref{verhulstsystem.eq}) or those embedded in figures \ref{fig4.fig}, \ref{fig7.fig} and \ref{fig13.fig}. The equation set includes accumulation of stocks, represented by the differential/flow  equations and a number of algebraic equations which represent the causal pathways between stocks. A system dynamics model is reduced to a differential equation representation by collapsing these causal pathways using substitution.  The causally connected differential equation representation provides a way of retaining the pathway information in differential equation form so that individual forces, and their impacts, can be identified. The method  retains pathway labels to distinguish pathways between stocks. The method labels the pathway with the names of the intermediary variables in the causal chain.

Consider an algebraic equation in functional form: $y=f(x)$, where $x$ is the cause of effect $y$. Rewrite the equation with the effect, the LHS of the equation,  as a subscript on the cause $y=f (x_{\underline{y}})$. The pathway between the two variables is now labelled by the effect; underlined to distinguish it from variable labels. Let $y$ be the cause of a further effect $z$: $z=g(y)$. Then the causal chain from $x$ to $z$ becomes $z=g(y) = g(f (x_{\underline{y}}))$. Although the value of $y$ has been eliminated by substitution, its name is retained as a subscript on the first cause $x$.

Further, if $z$ is a cause of $w$, $w=h(z) $, continuing the chain, then $w$ can now be written as a function of $x$ with the pathway through $y$ and $z$ retained. Using $z=g(y_{\underline{z}})$:
$$w = h(z) = h(g(y_{\underline{z}})) = h(g(f(x_{\underline{y}})_{\underline{z}})) = h(g(f(x_{\underline{yz}})))$$
where the definition $f(x_{\underline{y}})_{\underline{z}}  \triangleq f(x_{\underline{yz}})$ has been used.
The notation can be extended to functions with many arguments \cite{hayward2017newton}.

For example, in the Verhulst model (\ref{verhulstsystem.eq}):
\begin{eqnarray*}
f=1-\frac{x}{M} &\rightarrow& f=1-\frac{x_{\underline{f}}}{M} \\
g=rf &\rightarrow& g=rf_{\underline{g}} = r\left(1-\frac{x_{\underline{fg}}}{M}\right) \\
G=gx &\rightarrow& G=g_{\underline{G}} x_{\underline{G}} =r \left(1-\frac{x_{\underline{fgG}}}{M}\right) x_{\underline{G}}
\end{eqnarray*}
 Thus the two pathways from $x$ to its rate of change $\dot{x}=G$ are distinguished.
 
For an example with two variables, the four causal pathways to predator $y$, figure  \ref{fig13.fig}, are labelled:
 \begin{eqnarray*}
\gamma=cx &\rightarrow& \gamma = cx_{\underline{\gamma}}\\
F=\gamma y&\rightarrow& F=\gamma_{\underline{F}} y_{\underline{F}} = c x_{\underline{\gamma F}} y_{\underline{F}}\\
\delta = d+ey &\rightarrow& \delta = d+ey_{\underline{\delta}} \\
G=\delta y &\rightarrow&  G=\delta_{\underline{G}} y_{\underline{G}}=(d+e y_{\underline{\delta G}} )y_{\underline{G}}
\end{eqnarray*}
The differential equation (\ref{lv6.eq}) follows from $\dot{y}=F-G$. Once pathways have been distinguished by the intermediary variables, they can be renamed to reflect  either the force they represent or the associated feedback loop as is common in system dynamics.
 
\section {Linearised Impacts in the Predator-Prey Model} \label{appendix2}
The computation of the second-order impacts at equilibrium requires linearisation.   Let $x= 1+ae + \phi$ and $y = a + \psi$, where $\phi$ and $\psi$ are small. Thus, the linearised differential equations (\ref{lv1.eq}--\ref{lv2.eq}), with $b=c=d=1$, become:
\begin{eqnarray}
\dot{\phi} &=& -(1+ae) \psi \label{phieq.eq} \\
\dot{\psi} &=& a \phi  - ae \psi \label{psieq.eq}
\end{eqnarray}
In the case of real eigenvalues, the general solution of the equations are
\begin{eqnarray*}
\phi &=&   A\tfrac{( ae+\lambda_+)}{a} e^{t\lambda_+} + B \tfrac{(ae+\lambda_-)}{a} e^{t \lambda_-} \\
\psi &=& A e^{t\lambda_+} + B e^{t \lambda_-}
\end{eqnarray*}
where $A$ and $B$ are constants and $ \lambda_{\pm} =  -ae/2 \pm \sqrt{a^2e^2-4a-4a^2e}/2$ are the eigenvalues of the system matrix. Both eigenvalues are negative with the larger, $\lambda_+$, dominating as $t\rightarrow \infty$. Thus, the impact $\Ip_{\mathrm{pred}\,x}(B_4)$ (\ref{impactpp6.eq}) can be evaluated at equilibrium independently of the constants $A,B$:
$$
\Ip_{\mathrm{pred}\,x}(B_4) = - \frac{y(x - 1-ey) }{a  - y}  
 \approx    a\frac{A(\tfrac{\lambda_+ }{a}) e^{t\lambda_+} }{A e^{t\lambda_+} }
=\lambda_+  
= - \frac{ae - \sqrt{a^2e^2-4a-4a^2e}}{2} 
$$
The result for $\Ip_{\mathrm{pred}\,y}(B_4)=  -(1+ae) a/\lambda_+   $ follows in a similar manner.

In the case of complex eigenvalues, the imaginary part, $\omega = \sqrt{4a+4a^2e-a^2e^2}/2$, is the frequency of oscillation, and $-ae/2$ the damping coefficient. Thus, the solution of (\ref{phieq.eq}--\ref{psieq.eq}) is:
\begin{eqnarray*}
\phi&= &e^{-\frac{ae}{2}t} \left( P \sin(\omega t + \epsilon) + Q \cos(\omega t + \epsilon)    \right) \\
\psi &=&e^{-\frac{ae}{2}t} R \cos(\omega t + \epsilon)
\end{eqnarray*}
where $R$ and $\epsilon$ are constants with $ P = \omega/a R$ and $Q=eR/2$. Thus, the second-order impact on $x$ is:
$$
\Ip_{\mathrm{pred}\,x}(B_4) =  a \frac{-R e^{-\frac{ae}{2}t} \left( \frac{\omega}{a}  \sin(\omega t + \epsilon) + \frac{e}{2}  \cos(\omega t + \epsilon)        \right)}{ -e^{-\frac{ae}{2}t}  R \cos(\omega t + \epsilon)} 
= - \omega  \tan(\omega t + \epsilon) - \frac{ae}{2} 
$$

Although this impact does not have a limit as $t\rightarrow \infty$, it represents a ratio measure of a now infinitesimally small force on $x$ which can be replaced with its average over one cycle. The average of the impact over one cycle can be computed by taking limits: 
$$
\left<\Ip_{\mathrm{pred}\,x}(B_4) \right> = -  \lim_{\eta \rightarrow  \pi/2} \int^{\eta/\omega - \epsilon/\omega}_{-\eta/\omega - \epsilon/\omega}  \omega  \tan(\omega t + \epsilon) dt - \frac{ae}{2}   
 =  - \lim_{\eta \rightarrow  \pi/2} \left[ \ln \cos(\eta) - \ln \cos(-\eta) \right]    - \frac{ae}{2} = -\frac{ae}{2} 
$$
The impact of predation on $y$ can be computed in a similar manner $<\Ip_{\mathrm{pred}\,y}(B_4)>  =\frac{ae}{2} $. The product of the average impacts is not equal to the loop gain as the loop impact theorem of \citet{hayward2014model}
only applies at an instant in time where, in this case, the impacts of $x$ and $y$ are out of phase.

%
%
%
%
%
%
%
%

\end{document}